# Temperature-Dependent Neutron Moderation Model Including Inelastic Scattering in Reactor Media

**Sergey Chernezhenko[1], Victor Tarasov[2], Volodymyr Vashchenko[3], Iryna Korduba[4]**

[1] Odessa National Polytechnic University, Shevchenko Ave. 1, Odesa, 65044, Ukrainehttps://orcid.org/0000-0003-2840-9003
E-mail: chernezhenko@op.edu.ua
[2] Odessa National Polytechnic University, Shevchenko Ave. 1, Odesa, 65044, Ukraine
https://orcid.org/0000-0002-1497-0214
E-mail: tarasovva@op.edu.ua
[3] National Aviation University, Kyiv, 03000, Ukraine
https://orcid.org/0000-0003-1585-2129
E-mail: nucleoroid@gmail.com
[4] Kyiv National University of Construction and Architecture, 31, Povitroflotskyi ave., Kyiv, 03037, Ukraine,
https://orcid.org/0000-0001-5135-8465
E-mail: korduba.ib@knuba.edu.ua

**Abstracts**. In this study, a mathematical model of neutron moderation is developed that accounts for the temperature of the fissile medium and the contribution of inelastic neutron scattering on heavy nuclei (U-238) in reactor cores. Within the gas model framework, an analytical expression for the inelastic neutron scattering law for an isotropic neutron source is derived for the first time, incorporating the temperature of the moderating medium as a parameter. The scattering law is obtained from the general kinematic solution of inelastic neutron–nucleus interactions in the laboratory (“L”) system, where both particles possess arbitrary velocity vectors. Based on the newly derived inelastic and previously obtained elastic scattering laws, analytical formulas for the neutron flux density and slowing-down spectrum are presented, both dependent on the medium’s temperature. The calculated deceleration spectra exhibit two distinct maxima: a high-energy and a low-energy peak. The low-energy maximum is consistent with the analytical solution of the neutron balance equation, confirming the validity of the proposed model. The developed approach provides a deeper understanding of neutron energy distribution in temperature-dependent reactor media and can be applied to improve the accuracy of neutron-kinetic calculations in thermal and fast reactor systems.

## Introduction

An important link in nuclear reactor physics is the theory of neutron deceleration (Kamal, 2014, 2014; Marguet, 2018; Melnikov et al., 2015; Stacey, 2017). The traditional theory of neutron deceleration for nuclear reactor physics does not include the temperature of the decelerating medium because it is based on the law of neutron scattering, which does not take into account the thermal motion of the medium nuclei. Within the framework of this theory, the spectrum of slowing neutrons is described by an analytical expression called the Fermi-Wigner deceleration spectrum (Verkhivker and Kravchenko, 2008). The Fermi-Wigner deceleration spectrum also does not depend on the temperature of the decelerating medium and therefore cannot correctly describe the experimental spectrum of decelerating neutrons in its low-energy part, for example, the spectra of thermal nuclear reactors. Since the experimental neutron spectra of thermal nuclear reactors are similar to the Maxwell distribution (Stacey, 2017, 2010), it was assumed that the spectrum of slowing neutrons in its low-energy part can be described in the form of the Maxwell distribution. This assumption allowed us to develop a semi-experimental calculation scheme for obtaining the neutron deceleration spectra, which is widely used today in nuclear reactor physics. According to this calculation scheme, the decelerating neutron spectrum is a combination of the high-energy part in the form of the Fermi-Wigner deceleration spectrum and the low-energy part of the spectrum in the form of the Maxwell distribution. To set the low-energy part in the form of the Maxwell distribution, it is necessary to set the temperature of the neutron gas, for which the formula for recalculating the temperature of the decelerating medium to the temperature of the neutron gas is used (Marguet, 2023, 2018). Moreover, this conversion formula, according to (Melnikov et al., 2015), was obtained by numerical approximation of the experimental spectra of several thermal nuclear reactors of different types available at that time, and is still widely used in the physics of nuclear reactors, e.g.,(V. Rusov et al., 2011; Rusov et al., 2012; V. D. Rusov et al., 2011; VA Tarasov et al., 2023). Note also that this semi-experimental computational scheme for obtaining neutron deceleration spectra also includes a procedure for stitching together the high-energy part of the Fermi-Wigner deceleration spectrum and the low-energy part of the spectrum in the form of a Maxwell distribution, by means of which the neutron boundary energy between them is determined. The above considerations indicate that there is still no comprehensive theory of neutron slowing down that accounts for the thermal motion of nuclei in the moderating medium, emphasizing the need for its development.

It should also be noted that neutron slowing-down spectra in various media are frequently obtained using the Monte Carlo method. This approach forms the basis of modern computational tools for modeling neutron moderation and transport processes in reactor systems. For instance, software packages such as MCNP4 and GEANT4 provide user interfaces and simulation capabilities for calculating neutron and photon transport in complex reactor geometries using Monte Carlo techniques (Briesmeister, 2000; Rusov et al., 2015b) and others allow one to obtain the spectrum of decelerating neutrons by the Monte Carlo method by performing a computer calculation of the kinematic problem of neutron scattering on the nucleus (taking into account possible neutron absorption for decelerating media, absorbing neutrons) a huge number of times, setting for each calculation the initial velocities of the neutron and the nucleus with the help of random counters, and obtaining the resulting spectrum of decelerating neutrons by averaging the accumulated results of calculations for all realizations. Clearly, it is difficult to call this yield a theory.

The lack of a theory of neutron deceleration, first of all, creates problems in the study of emergency modes at nuclear reactors and in the development of the physics of new generation reactors, e.g., nuclear fission traveling wave reactors (Rusov et al., 2015a; Shcherbyna et al., 2024; Viktor Tarasov et al., 2023), pulsed reactors, neutron multiplier reactors (boosters), subcritical assemblies (Hayes, 2017; Shcherbyna et al., 2024; Tilak and Basu, 2015), as well as, natural nuclear reactors, e.g., georeactor (Rusov et al., 2007).

The paper develops a basic model of the neutron deceleration theory, which takes into account the temperature of the fissile medium and has been recently published in our paper (Tarasov, 2017). The basic model of the neutron deceleration theory does not take into account the contribution of the inelastic neutron scattering reaction on heavy nuclei of reactor fissile media, and therefore it is applicable to decelerating media consisting of nuclei for which the inelastic scattering reaction cross sections are negligibly small. In this paper, we develop a theory of the slowing down of neutrons emitted by an isotropic neutron source that takes into account the contribution of inelastic neutron scattering reactions on heavy nuclei of reactor fissile media. Accounting for inelastic scattering reactions is especially important for moderating reactor fissile media consisting of heavy nuclei. According to, for example, inelastic neutron scattering on heavy nuclei will be observed at neutron energies exceeding several hundred kiloelectronvolts, and on light nuclei - at energies exceeding one or several megaelectronvolts.

The field of scientific interests of the authors is related to the physics of fissile neutron multiplying media (reactor fuel media) and therefore the theory of neutron slowing down was developed by the authors in such a way that it was first of all applicable to such media. The assumption of neutron source isotropy is consistent with the properties of neutron source density in fissile neutron multiplying media (Kamal, 2014; Marguet, 2023; Melnikov et al., 2015; Stacey, 2017).

The reactor fissile (fuel) medium is in thermodynamic nonequilibrium during reactor operation (and, in general, any fissile medium in which the process of chain fissions is initiated in one way or another), since it undergoes fission processes accompanied by the release of a large amount of energy, emission of neutrons and other particles, changes in nuclide composition, heat transfer, dynamics of radiation defects leading to changes in the geometrical parameters of the medium and even to its destruction, etc. The reactor fissile (fuel) medium is in a thermodynamic nonequilibrium state. In other words, the reactor fissile medium in which fission processes take place is an open physical system in a thermodynamic non-equilibrium state. Such a system is described within the framework of non-linear, non-equilibrium thermodynamics of open physical systems. In such systems, non-equilibrium stationary states may exist, which satisfy the Prigogine criterion: the minimum of entropy production (Ben-Naim and Casadei, 2017; Glansdorff and Prigogine, 1970). As is known from nonlinear nonequilibrium thermodynamics, the realization and type of such a stationary regime depends not only on the internal parameters of the system (internal entropy), but also on the boundary conditions (entropy fluxes at the boundary), e.g., for the realization of a stationary state in a nonequilibrium system (the so-called nonequilibrium-stationary state) it is necessary to keep the boundary conditions constant (Glansdorff and Prigogine, 1970).

In this work, the following simplifications are adopted in constructing a theoretical model of the neutron deceleration process in a fissile medium. Two thermodynamic subsystems are distinguished in a fissile medium: the subsystem of decelerating neutrons and the subsystem of decelerating medium nuclei. Each of them is an open physical system, and they interact with each other. According to the above mentioned in the previous paragraph, in reality both of these systems are in a nonequilibrium state. However, in our model, we adopt the simplification that the subsystem of the moderating medium nuclei is in a state close to its equilibrium state because of its inertia with respect to perturbations and, therefore, neglecting the influence of the neutron subsystem on it, we assume that the subsystem of the moderating medium nuclei is in thermodynamic quasi-equilibrium. This simplifying assumption allows us to introduce the temperature of the decelerating medium into the model and describe the reshuffling of the energies of kinetic motion of the decelerating medium nuclei by the Maxwell distribution. We consider the subsystem of decelerating neutrons as nonequilibrium and do not introduce the temperature for it in our model.

We emphasize that, as noted above, the temperature of the neutron gas subsystem is introduced in the traditional approach to construct the neutron deceleration spectrum. This indicates the existence in it of a simplifying assumption, which we have abandoned, that the neutron subsystem is also in a state of thermodynamic equilibrium. Moreover, as indicated above, the neutron gas temperature formula is related to the temperature of the moderating medium by means of a relation empirically fitted to the experiment.

In this paper, in the framework of the gas model, for the first time an analytical expression for the inelastic neutron scattering law for neutrons interacting with the nuclei of the medium through the inelastic neutron scattering reaction channel, is obtained for an isotropic neutron source, including the temperature of the moderating medium as a parameter. The obtained scattering law is based on the solution of the kinematic problem of inelastic scattering of a neutron on a nucleus in the “L”-system in the general case, i.e., when before scattering not only the neutron but also the nucleus possesses an arbitrarily given velocity vector in the “L”-system. For the elastic scattering law obtained by the authors earlier [20] and the new inelastic scattering law, analytical expressions for the flux density and the neutron deceleration spectrum in the reactor fissile medium have been found, taking into account elastic and inelastic scattering and depending on the temperature of the medium. The obtained expressions for the spectra of decelerating neutrons allow a new interpretation of the physical nature of the processes that determine the neutron spectrum in the thermal region.

The purpose of this study was to derive analytical expressions for the law of inelastic neutron scattering, neutron flux density, and the neutron retardation spectrum for an isotropic neutron source in various retarding neutron-absorbing media, with the temperature of the retarding medium included as a critical parameter, and to validate these models through comparative analysis with computational tools like GEANT4 (Chadwick et al., 2011) and the Monte Carlo method (Pollard, 1978). The research aimed to deepen the understanding of the physical processes shaping neutron spectra in the low-energy region by exploring the impact of elastic and inelastic neutron scattering cross-sections and the logarithmic energy decrement on the spectrum's maximum. The scientific novelty lies in obtaining these analytical expressions for the first time, offering a new interpretation that links the maximum of the neutron retardation spectrum to the interaction of a non-equilibrium neutron system with a thermalized nuclear system, rather than solely the thermodynamically equilibrium part described by the Maxwell distribution. This departs significantly from traditional models and provides a foundation for experimental verification. Additionally, the study highlights the potential for investigating the diverse behavior of scattering cross-sections across different retarding media, opening new avenues for nuclear technology advancements.

**Methods**

*The Monte Carlo method* is a numerical statistical technique based on the use of random number generators to solve problems with inherently probabilistic processes (Pollard, 1978). In reactor physics, this method is widely applied to simulate neutron transport in various media, particularly to determine neutron slowing-down spectra while accounting for scattering and absorption reactions.

The core idea of the method involves repeated simulation of individual neutron trajectories, where each simulation includes the generation of initial parameters (such as energy, direction of motion, and the velocity of medium nuclei) using random variables, determining the probability of different interaction types (elastic or inelastic scattering), calculating new parameters after collisions, and accumulating statistical data on the energy distribution of neutrons over a large number of realizations.

The advantage of the Monte Carlo method lies in its high accuracy in reproducing physical processes, provided a sufficient number of simulations is conducted. However, its drawbacks include high computational demands and the need for significant statistical sampling to achieve reliable results. The method is implemented in modern software packages such as MCNP, GEANT4, and Serpent, which are widely used in nuclear engineering and reactor physics.

*GEANT4 as a Tool for Simulating Particle-Matter Interactions*

GEANT4 (GEometry ANd Tracking) is an open-source Monte Carlo–based software toolkit developed under the auspices of CERN for simulating the transport of elementary particles and their interactions with matter over a wide energy range, from electronvolts (eV) to teraelectronvolts (TeV) (Mendoza et al., 2014; Sadeq et al., 2022). It is widely used in high-energy physics, nuclear physics, medical physics, space science, and radiation protection.

GEANT4 (Agostinelli et al., 2003) includes comprehensive models for all major classes of particle interactions with matter, including electromagnetic, hadronic (strong), and weak interactions, as well as nuclear processes such as elastic and inelastic scattering, absorption, secondary particle production, and radioactive decay. Its modular architecture allows users to tailor the simulation to specific problems by defining the geometry of the system, material properties, particle sources, and the physical models used.

In the context of nuclear reactor physics and neutron slowing-down analysis, GEANT4 enables accurate simulation of neutron transport and interactions with nuclei of the moderating medium. It accounts for medium temperature, anisotropy, energy losses, and secondary particle generation, making it a powerful tool for verifying analytical models and producing high-precision neutron slowing-down spectra in complex media.

For thermal and fast neutron applications, GEANT4 offers specialized physics lists such as QGSP_BIC_HP and QGSP_BERT_HP, which incorporate evaluated nuclear data libraries (G4NDL, based on ENDF/B-VII, JEFF, and JENDL) for accurate modeling of low-energy neutron interactions, including temperature-dependent cross sections and resonance behavior (Rhee et al., 2012; Sood, n.d.).

In nuclear engineering, GEANT4 is used to model processes such as neutron moderation, thermalization, diffusion, and capture in both critical and subcritical assemblies, shielding structures, and advanced reactor concepts. Its ability to simulate the kinematics of neutron scattering while considering thermal motion of nuclei allows for detailed studies of neutron energy spectra and spatial distribution, particularly important in the design and safety analysis of nuclear systems.

## Results and Discussion

### Kinematics of inelastic neutron scattering on the nucleus of a moderating medium

If the inelastic neutron scattering reaction is characteristic for the nuclei of a neutron-decelerating medium, it is necessary to clarify the kinematics presented in our work (Tarasov, 2017).

According to (Kamal, 2014; Marguet, 2018), the inelastic neutron scattering reaction on a nucleus proceeds through the stage of formation of a compound nucleus in an excited state, i.e., at the first stage of the reaction, the nucleus of the moderating medium captures a neutron:

$$(\mathrm{A},\mathrm{Z}) + {}_0^1n \rightarrow (A+1,Z)^* \quad (1)$$

where: $(\mathrm{A},\mathrm{Z})$- nucleus of the moderating medium with mass number A and charge Z; ${}_0^1n$ - neutron (0 - zero charge, 1 - neutron mass number); $(A+1,Z)^*$ - composite nucleus with mass number $(\mathrm{A}+1)$and charge Z; $*$- designation of the excited state of the nucleus.

At the next stage of the reaction the compound nucleus decays with formation of the moderating medium nucleus$(\mathrm{A},\mathrm{Z})^*$(but in excited state) and emission of neutron, and at the final stage of the reaction the excitation of the nucleus$(\mathrm{A},\mathrm{Z})^*$ is removed by emission of $\gamma$ - quantum:

$$(A+1,Z)^* \rightarrow (A,Z)^* + {}_0^1n \rightarrow (A,Z) + {}_0^1n + \gamma \quad (2)$$

Here it should be noted that the decay of the composite nucleus with formation of the decelerating medium nucleus$(A,Z)^*$ and neutron emission is a two-particle nuclear reaction, and the kinematic problem in this case is the problem of two-particle inelastic scattering, which is a special case of the kinematic problem for an endothermic nuclear reaction. In any nuclear reaction, the law of conservation of total momentum and the law of conservation of total energy are satisfied, and the total energy of each particle is equal to the sum of its internal energy and its kinetic energy. In an endothermic nuclear reaction, the total internal energy of the particles after the reaction is as follows$\varepsilon$is greater than the total internal energy of particles before the reaction$\varepsilon_0$, that is, the so-called heat effect of the reaction $Q = \varepsilon_0 - \varepsilon < 0$. It follows from the law of conservation of total energy that also $Q = T - T_0$where $T$ and $T_0$are the total kinetic energy after the reaction and before the reaction, respectively. Thus the problem of inelastic scattering is a problem in which the law of conservation of total momentum is satisfied and the law of conservation of kinetic energy is violated, but the violation of the law of conservation of kinetic energy can be taken into account if the thermal effect of the reaction $Q = T - T_0$is known. For us it is important that the kinematical problem of the two-particle endothermic reaction is solved uniquely (Sitenko and Malnev, 1994). And, consequently, the kinematical problem of two-particle inelastic scattering is also solved uniquely. Indeed, let us represent the inelastic neutron scattering reaction by restricting ourselves to the stage of decay of the composite nucleus with neutron emission:

$$(\mathrm{A},\mathrm{Z}) + {}_0^1n \rightarrow (A+1,Z)^* \rightarrow (\mathrm{A},\mathrm{Z})^* + {}_0^1n \quad (3)$$

For comparison, we also give a typical calculation scheme of a two-particle inelastic scattering reaction on the example of a nucleus $(\mathrm{A},\mathrm{Z})$ and the neutron${}_0^1n$.

$$(\mathrm{A},\mathrm{Z}) + {}_0^1n \rightarrow (\mathrm{A},\mathrm{Z})^* + {}_0^1n \quad (4)$$

Consequently, reactions (3) and (4) differ only by the existence of an intermediate nucleus, which causes different reaction times. However, the laws of conservation of total momentum and total energy for both reactions (3) and (4) are the same. Therefore, the solutions of the kinematic problem for reactions (3) and (4) will be the same.

Thus, one can find the kinematics of a neutron interacting with the nucleus through the inelastic scattering reaction channel by solving the kinematic problem of two-particle inelastic scattering according to the reaction scheme (4).

Let us consider the kinematics of inelastic scattering of a neutron on a nucleus moderating medium. The neutron-decelerating medium, as in (Tarasov, 2017), is described in the framework of the gas model, i.e., it is assumed that the nuclei of the medium have kinetic energy due to their thermal motion. As in (Tarasov, 2017), we introduce two laboratory systems (Fig. 1):

- a resting laboratory coordinate system, which we will call the laboratory coordinate system "L";

- the laboratory coordinate system "L'" moving relative to the laboratory coordinate system "L" with a constant velocity equal to the thermal motion velocity of the decelerating medium nucleus on which the neutron scattering occurs.

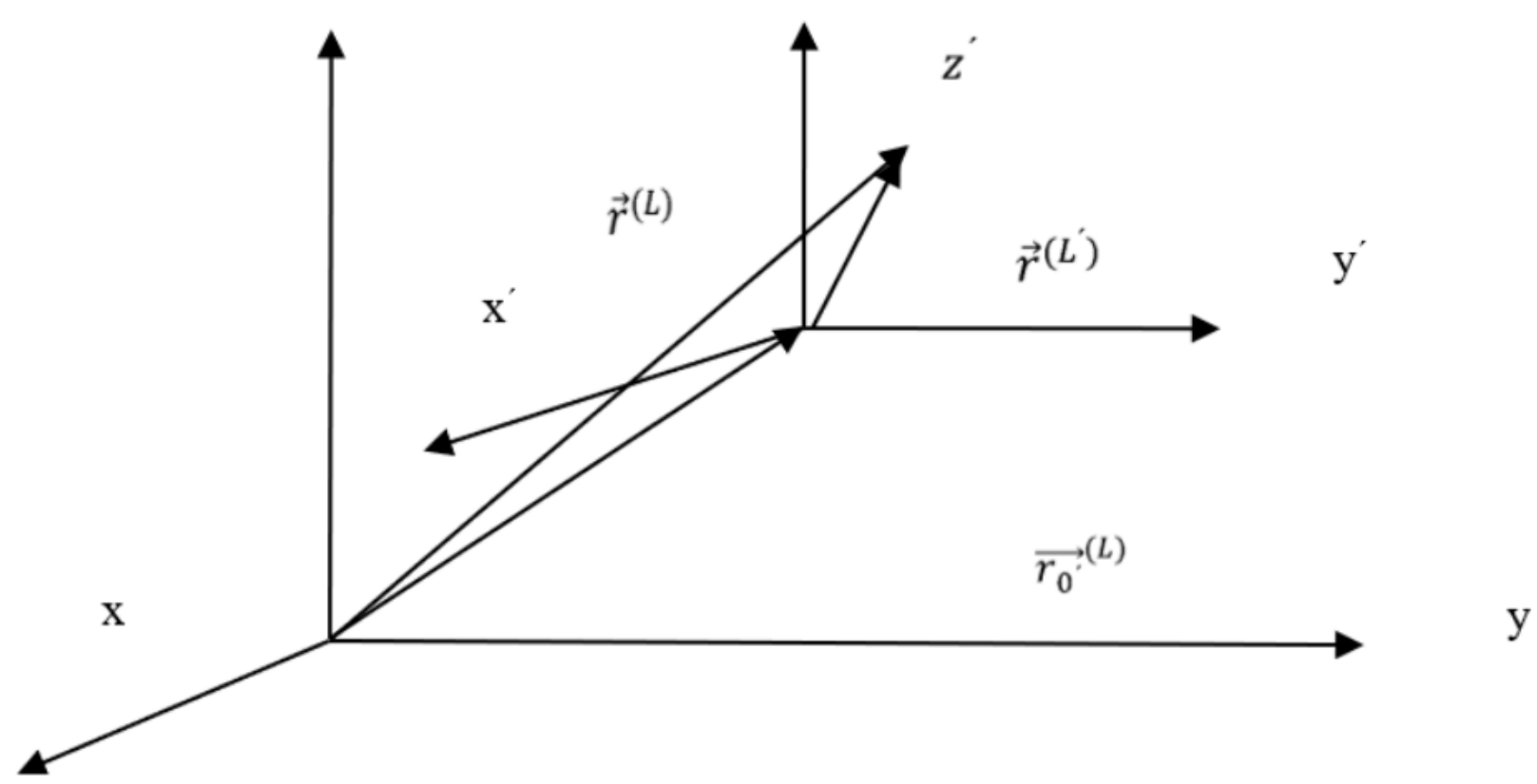

*Fig. 1.* Laboratory coordinate systems "L" and "L'".

Note that we consider a special case when the spatial orientation of the coordinate axes of the laboratory coordinate systems "L" and "L'" is the same and the radius vector of the origin of the laboratory coordinate system "L'" in the laboratory coordinate system "L" coincides with the radius vector of the decelerating medium nucleus, on which the neutron scattering occurs, in the laboratory coordinate system "L", i.e., the decelerating medium nucleus in the laboratory coordinate system "L'" is at rest.

We use the same notations as in [20]:$m_1 = m_n$ - neutron mass; $m_1 = m_L$ - the mass of the nucleus; $\overrightarrow{r_1}^{(L)}$– radius-vector of the neutron in the laboratory system "L";$\overrightarrow{r_2}^{(L)}$ radius-vector of the nucleus in the laboratory system "L"; $\overrightarrow{r_{\text{Ts}}}^{(L)}$- radius- vector of the center of inertia in the laboratory system "L";$\overrightarrow{r_1}^{(L')}$ - radius-vector of the neutron in the laboratory system "L'";$\overrightarrow{r_2}^{(L')}$ - - radius-vector of the nucleus in the laboratory system "L'";$\overrightarrow{V_{10}}^{(L)}$ - neutron velocity in the "L" system before collision with the nucleus; $\overrightarrow{V_1}^{(L)}$- neutron velocity in the "L" system after collision with the nucleus; $\overrightarrow{V_{20}}^{(L)}$- velocity of the nucleus in the "L" system before collision with the neutron; - velocity of the nucleus in the "L" system after collision with the neutron; $\overrightarrow{V_{10}}^{(L')}$ - neutron velocity in the "L'" system before collision with the nucleus; $\overrightarrow{V_1}^{(L')}$ - neutron velocity in the "L'" system after collision with the nucleus; $\overrightarrow{V_{20}}^{(L')}$ - the velocity of the nucleus in the “L'” system before the collision with the neutron; $\overrightarrow{V_2}^{(L')}$ - velocity of the nucleus in the “L'” system after collision with a neutron; $\overrightarrow{V_{Ts}}^{(L')}$ - velocity of the center of inertia in the laboratory system “L'”.

The radius vectors of a point in the laboratory systems “L” and “L'” are related by the following equality:

$$\vec{r}^{(L)} = \overrightarrow{r_0}^{(L)} + \vec{r}^{(L')} \tag{5}$$

Thus, in accordance with the task we can write that:

$$\overrightarrow{V_{10}}^{(L)} = \frac{d\vec{r_1}^{(L)}}{dt} \neq 0 \text{and } \overrightarrow{V_{20}}^{(L)} = \frac{d\vec{r_2}^{(L)}}{dt} \neq 0 \tag{6}$$

In the laboratory coordinate system “L'” the nucleus is at rest before the collision with the neutron, i.e.:

$$\overrightarrow{V_{20}}^{(L')} = 0 \tag{7}$$

Then the connection between coordinates m1 and m2 in coordinate system “L” and coordinate system “L'” is given by expression (5), and between velocities by the following expressions (the law of addition of non-relativistic velocities for inertial systems, following from the principle of relativity of Galileo):

$$\overrightarrow{V_1}^{(L')} = \overrightarrow{V_1}^{(L)} - \overrightarrow{V_{20}}^{(L)} \text{ and} \overrightarrow{V_2}^{(L')} = \overrightarrow{V_2}^{(L)} - \overrightarrow{V_{20}}^{(L)} \tag{8}$$

The solution of the kinematic problem of inelastic scattering of a neutron on a nucleus in the laboratory coordinate system “L'” can be written in the form analogous to the solution forkinematics of a two-particle endothermic nuclear reaction with the thermal effect of the reaction Q different from zero. Indeed, according to the scheme of the inelastic neutron scattering reaction (4), the thermal effect of this reaction is as follows $|Q|$ is equal to the energy of the emitted γ - quantum, which we denote by $E_\gamma$ and then let us consider our problem similarly to the problem in (Sitenko and Malnev, 1994), but in our previously introduced notations.

The problem of the collision of two particles in the laboratory coordinate system “L'” is convenient to solve in the system of the center of inertia, i.e., in the “Ts” system.

For the reaction of inelastic scattering of a neutron in the “Ts” system, the law of conservation of total momentum of a closed system is satisfied, but the law of conservation of total kinetic energy of this closed system is not satisfied.

Proceeding from the law of conservation of momentum, for two colliding particles in the “CD” system we obtain:

$$\overrightarrow{P_{10}}^{(Ts)} + \overrightarrow{P_{20}}^{(Ts)} = \overrightarrow{P_1}^{(Ts)} + \overrightarrow{P_2}^{(Ts)} = 0 \tag{9}$$

where: $\overrightarrow{P_{20}}^{(Ts)}$ - impulses of the neutron and the nucleus in the CD system before the interaction, respectively;

$\overrightarrow{P_1}^{(Ts)}$ and $\overrightarrow{P_2}^{(Ts)}$ - are the momenta of the neutron and the nucleus in the CD system after the interaction. From relations (9) it follows:

$$m_1 \cdot \overrightarrow{V_{10}}^{(Ts)} = m_2 \cdot \overrightarrow{V_{20}}^{(Ts)} \text{and} m_1 \cdot \overrightarrow{V_1}^{(Ts)} = m_2 \cdot \overrightarrow{V_2}^{(Ts)} \tag{10}$$

From formulas (10) for the velocity moduli we obtain:

$$\left|\overrightarrow{V_{10}}^{(Ts)}\right| = \overrightarrow{V_{10}}^{(Ts)} = \frac{m_2}{m_1}\left|\overrightarrow{V_{20}}^{(Ts)}\right| = \frac{m_2}{m_1}\overrightarrow{V_{20}}^{(Ts)} \text{and} V_1^{(Ts)} = \frac{m_2}{m_1} V_2^{(Ts)} \tag{11}$$

Introducing the mass number for the neutron and the nucleus $A_n = 1$ and $A_y = A$respectively, and assuming that $m_1 = m_n \approx A_n = 1$and $m_1 = m_y \approx A_n = 1$ for formulas (10) and (11) we obtain the following expressions:

$$\overrightarrow{V_{10}}^{(Ts)} = -A \cdot \overrightarrow{V_{10}}^{(Ts)} \text{and} \overrightarrow{V_{10}}^{(Ts)} = -A \cdot \overrightarrow{V_2}^{(Ts)} \tag{12}$$

and

$$\overrightarrow{V_{10}}^{(Ts)} = +A \cdot \overrightarrow{V_{20}}^{(Ts)} \text{ and } \overrightarrow{V_1}^{(Ts)} = +A \cdot \overrightarrow{V_2}^{(Ts)} \quad (13)$$

The coordinate of the center of inertia of the system of neutron and nucleus slowing down the medium, $\overrightarrow{r_{Ts}}^{(L')}$ - radius-vector of the center of inertia in the laboratory coordinate system "L'", can be written as follows:

$$\overrightarrow{r_{Ts}}^{(L')} = (1 \cdot \overrightarrow{r_1}^{(L')} + A \cdot \overrightarrow{r_2}^{(L')}) \cdot \frac{1}{A+1} \quad (14)$$

where: $\overrightarrow{r_1}^{(L')}$ - radius-vector of the neutron in the laboratory coordinate system "L'"; $\overrightarrow{r_2}^{(L')}$ - radius-vector of the nucleus.

Taking into account that in the laboratory coordinate system "L'" the velocity of the nucleus before the collision $\overrightarrow{V_{20}}^{(L')} = 0$the velocity of the center of inertia for a closed system of two particles (here a neutron and a nucleus) in the laboratory coordinate system "L'" according to (14) is written as follows:

$$\overrightarrow{V_{Ts}}^{(L')} = \frac{1}{A+1} \cdot \overrightarrow{V_{10}}^{(L')} \quad (15)$$

By virtue of the law of conservation of total momentum, the velocity of the center of inertia in the system "L'" before and after the collision will not change, and therefore the indices corresponding to the values before the interaction and after the interaction are omitted for the velocity of the center of inertia.

Since the system of the center of inertia (system "Ts") moves relative to the laboratory system with the velocity of the center of inertia in system "L'", then for the velocity of the neutron in system "Ts" before the interaction we have:

$$\overrightarrow{V_{10}}^{(Ts)} = \overrightarrow{V_{10}}^{(L')} - \overrightarrow{V_{Ts}}^{(L')} \quad (16)$$

then we substitute expression (15) into this expression and obtain:

$$\overrightarrow{V_{10}}^{(Ts)} = \overrightarrow{V_{10}}^{(L')} - \frac{1}{A+1} \cdot \overrightarrow{V_{10}}^{(L')} = \frac{A}{A+1} \cdot \overrightarrow{V_{10}}^{(L')} \quad (17)$$

Using the equations (10) or expression (15), taking into account that $\overrightarrow{V_{20}}^{(L')} = 0$ we find the velocity of the nucleus in the "C" system before the collision:

$$\overrightarrow{V_{20}}^{(Ts)} = \overrightarrow{V_{20}}^{(L')} - \frac{1}{A+1} \cdot \overrightarrow{V_{10}}^{(L')} = -\frac{1}{A+1} \cdot \overrightarrow{V_{10}}^{(L')} \quad (18)$$

The total kinetic energy of our system after the reaction (a system of two particles $(A, Z)^*$ and ${}_0^1n)T^{Ts}$ is equal to:

$$T^{Ts} = T_0^{Ts} + |Q| = T_0^{Ts} + E_\gamma \quad (19)$$

where: $T_0^{Ts}$- is the total kinetic energy of our system before the reaction.

For the total kinetic energy of our system before the reaction, using expressions (17) and (18), we obtain the expression:

$$T_0^{Ts} = \frac{\left|\overrightarrow{P_{10}}^{(Ts)}\right|^2}{2\mu} = \frac{\mu\left(\overrightarrow{V_{10}}^{(L')}\right)^2}{2} \quad (20)$$

where: μ - is the reduced mass of the particles of the system before the reaction, which is equal to:

$$\mu = \frac{A \cdot 1}{A + 1}$$

Then the kinetic energy of our system after the reaction (4) is equal to:

$$T^{Ts} = \frac{\left|\overrightarrow{P_1}^{(Ts)}\right|^2}{2\mu} = T_0^{Ts} + E_\gamma = \frac{\left|\overrightarrow{P_{10}}^{(Ts)}\right|^2}{2\mu} + E_\gamma = \frac{\mu\left(\overrightarrow{V_{10}}^{(L')}\right)^2}{2} + E_\gamma \quad (21)$$

From (21) we obtain an expression for the neutron momentum modulus after the reaction:

$$\overrightarrow{P_1}^{(Ts)} = \sqrt{2\mu' T^{Ts}} = \sqrt{2\mu'(T^{Ts} + E_\gamma} = \sqrt{2\mu' \left[\frac{\mu \overrightarrow{V_{10}}^{(L')}}{2} + E_\gamma\right]} \quad (22)$$

where: $\mu'$ - is the reduced mass of particles of the system after the reaction, which is equal to: $\mu' = \frac{(A+E_\gamma/c^2)\cdot 1}{(A+E_\gamma/c^2)+1}$, where c is the speed of light in vacuum.

If we denote through the unit vector $\vec{e}_1$ directed along the neutron velocity in the CD-system after the reaction, then according to (22) we obtain:

$$\overrightarrow{V_1}^{(Ts)} = \frac{\left|\overrightarrow{P_1}^{(Ts)}\right|}{1} \cdot \vec{e}_1 = \sqrt{2\mu' \left[\frac{\mu \overrightarrow{V_{10}}^{(L')}}{2} + E_\gamma\right]} \cdot \vec{e}_1 \quad (23)$$

Note that the stochastic angular distribution of a large number of inelastically scattered neutrons, as well as for elastically scattered neutrons, can be considered spherically symmetric (Kamal, 2014; Marguet, 2018). Indeed, the neutron is emitted at the decay stage of the intermediate nucleus according to (2) and (3), and as is well known the lifetime of the composite nucleus is ~10-13 sec, which greatly exceeds the elastic scattering times ~10-20 sec. During the lifetime of a composite nucleus, the composite nucleus seems to forget the kinematic details of its formation and its decay is explained by a stochastic redistribution of energy between nucleons, as a result of which one neutron receives energy sufficient to escape from the nucleus (Kamal, 2014; Marguet, 2018). Therefore, the neutron formed during the decay of a composite nucleus can have any momentum direction independent of the initial conditions of the formation of the composite nucleus. Thus, the spatial distribution of momentum directions of neutrons escaping from composite nuclei formed in inelastic neutron scattering reactions is isotropic.

We obtain the vector of the neutron velocity after the reaction in the system "L'" if we add to the vector of the neutron velocity after the collision in the CD-system, given by expression (23), the vector of the CD-system velocity given by expression (15):

$$\overrightarrow{V_1}^{(L')} = \overrightarrow{V_1}^{(Ts)} + \frac{1}{A+1}\overrightarrow{V_{10}}^{(L')} \quad (24)$$

From the velocity parallelogram shown in (Fig. 2), we find the square of the modulus of the neutron velocity in the system "L'" after the collision:

$$\left(\overrightarrow{V_1}^{(L')}\right)^2 = \left(\overrightarrow{V_1}^{(Ts')}\right)^2 + \left(\overrightarrow{V_1}^{(L')} \cdot \frac{1}{A+1}\right)^2 + \frac{2\cdot(V_1^{(Ts)})\cdot(V_{10}^{(L')})}{A+1} \cdot \cos\theta \quad (25)$$

where: $\theta$- is the angle of departure of the neutron in the system of the center of inertia "Ts" (Fig. 2), counted from the direction of the neutron velocity vector before the interaction in the "L'" system. Indeed, according to (15), the direction of the vector $\overrightarrow{V_{Ts}}^{(L')}$ (Fig. 2) coincides with the direction of the vector $\overrightarrow{V_{10}}^{(L')}$.

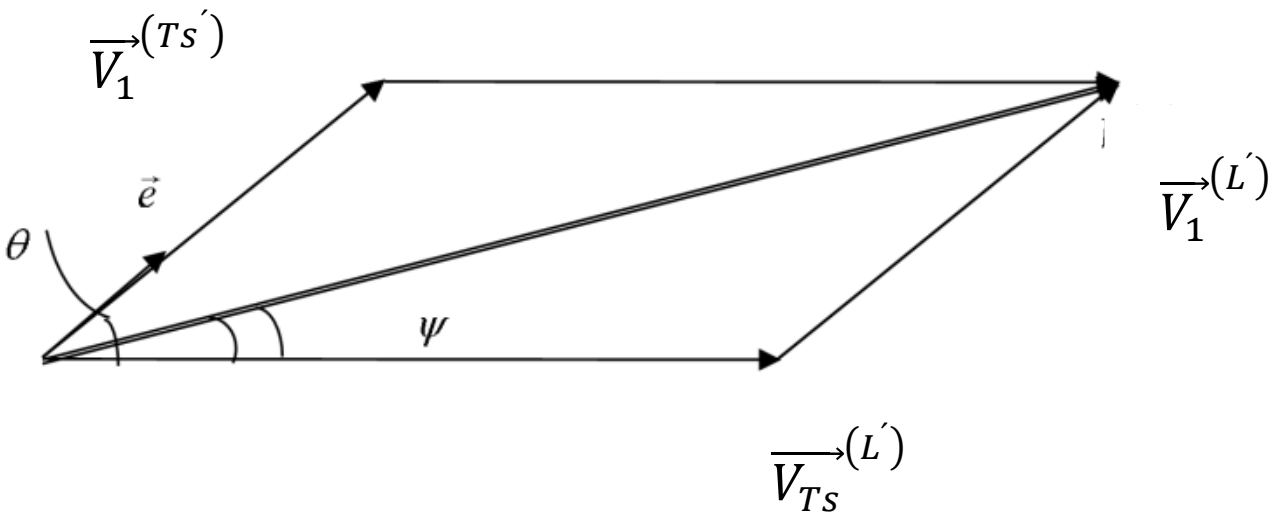

*Fig. 2.* Parallelogram of velocities after impact in the laboratory L' system.

Substituting in relation (25) the expression (15) for$\overrightarrow{V_{Ts}}^{(L')}$we can find the ratio of the square of the neutron velocity after inelastic scattering$\left(\overrightarrow{V_1}^{(L')}\right)^2$ and the square of the neutron velocity before scattering$\left(\overrightarrow{V_{10}}^{(L')}\right)^2$ in the laboratory coordinate system "L'", which is also equal to the ratio of the neutron kinetic energy after inelastic scattering$E_2^{(L')}$ and kinetic energy of the neutron before scattering $E_1^{(L')}$.

$$\frac{\left(\overrightarrow{V_1}^{(L')}\right)^2}{\left(\overrightarrow{V_{10}}^{(L')}\right)^2} = \frac{E_2^{(L')}}{E_1^{(L')}} = \frac{A^2+2A\cos\theta}{(A+1)^2} \tag{26}$$

If we introduce the inelastic scattering coefficient $\tilde{\alpha}$ as follows:

$$\tilde{\alpha} = \left(\frac{A-1}{A+1}\right)^2 \tag{27}$$

then expression (26) can be transformed to the following form, e.g., [5 - 7]:

$$\frac{E_2^{(L')}}{E_1^{(L')}} = \frac{1}{2}[(1+\tilde{\alpha}) + (1-\tilde{\alpha})\cos\theta] \tag{28}$$

Now, having expressed the vectors of neutron velocity before and after inelastic scattering in the "L'"-system through the analogous vectors of neutron velocity in the L-system according to expression (8), we can write expression (28) in the form of the following expression:

$$\frac{\left(\overrightarrow{V_1}^{(L)} - \overrightarrow{V_{20}}^{(L)}\right)^2}{\left(\overrightarrow{V_{10}}^{(L)} - \overrightarrow{V_{20}}^{(L)}\right)^2} = \frac{E_2^{(L')}}{E_1^{(L')}} = \frac{1}{2}[(1+\tilde{\alpha}) + (1-\tilde{\alpha})\cos\theta] \tag{29}$$

From expression (29) for inelastic scattering we obtain an expression similar to the expression for elastic scattering obtained in [20]:

$$\frac{\left(\overrightarrow{V_1}^{(L)}\right)^2}{\left(\overrightarrow{V_{10}}^{(L)}\right)^2} = \frac{E_2^{(L')}}{E_1^{(L')}} = \frac{1}{2}[(1+\tilde{\alpha}) + (1-\tilde{\alpha})\cos\theta]\left[1 - 2\frac{V_{10}^{(L)}V_{20}^{(L)}\cos\beta}{\left(\overrightarrow{V_{10}}^{(L)}\right)^2} + \frac{\left(\overrightarrow{V_{20}}^{(L)}\right)^2}{\left(\overrightarrow{V_{10}}^{(L)}\right)^2}\right] +$$

$$2\frac{\overrightarrow{V_1}^{(L)},\overrightarrow{V_{20}}^{(L)}}{\left(\overrightarrow{V_{10}}^{(L)}\right)^2} - \frac{\left(\overrightarrow{V_{20}}^{(L)}\right)^2}{\left(\overrightarrow{V_{10}}^{(L)}\right)^2} = \frac{1}{2}[(1+\tilde{\alpha}) + (1-\tilde{\alpha})\cos\theta] - \frac{1}{2}[(1+\tilde{\alpha}) + (1-$$

$$\tilde{\alpha})\cos\theta]\frac{V_{10}^{(L)}V_{20}^{(L)}\cos\beta}{\left(\overrightarrow{V_{10}}^{(L)}\right)^2} + \frac{V_1^{(L)}V_{20}^{(L)}\cos\gamma}{\left(\overrightarrow{V_{10}}^{(L)}\right)^2} - \left\{1 - \frac{1}{2}[(1+\tilde{\alpha}) + (1-\tilde{\alpha})\cos\theta]\right\}\cdot\frac{E_{20}^{(L)}}{E_{10}^{(L)}} \tag{30}$$

where: $\cos\beta$- cosine of the angle between vectors;

$\overrightarrow{V_{20}}^{(L)}$ which is expressed through the scalar product of these vectors $\left(\overrightarrow{V_{10}}^{(L)},\overrightarrow{V_{20}}^{(L)}\right)$ as follows:

$$\cos\beta = \frac{\overrightarrow{V_{10}}^{(L)},\overrightarrow{V_{20}}^{(L)}}{\left|\overrightarrow{V_{10}}^{(L)}\right|,\left|\overrightarrow{V_{20}}^{(L)}\right|} = \frac{\overrightarrow{V_{10}}^{(L)},\overrightarrow{V_{20}}^{(L)}}{\overrightarrow{V_{10}}^{(L)}\overrightarrow{V_{20}}^{(L)}} \tag{31}$$

$\cos\gamma$ - cosine of the angle between vectors $\overrightarrow{V_1}^{(L)}$ and $\overrightarrow{V_{20}}^{(L)}$.

$$\cos\gamma = \frac{\overrightarrow{V_1}^{(L)},\overrightarrow{V_{20}}^{(L)}}{\left|\overrightarrow{V_1}^{(L)}\right|,\left|\overrightarrow{V_{20}}^{(L)}\right|} = \frac{\overrightarrow{V_1}^{(L)},\overrightarrow{V_{20}}^{(L)}}{\overrightarrow{V_1}^{(L)}\overrightarrow{V_{20}}^{(L)}} \tag{32}$$

For our purposes, as we shall see below, we can restrict ourselves to expression (30) and not to give in the paper further transformations of expression (30) and related expressions (31) and (32) leading to the final set of expressions giving the exact solution of the kinematic problem of inelastic scattering of the neutron on the nucleus under consideration, taking into account the thermal motion of the nucleus, since this leads only to a set of cumbersome expressions. This is due to the fact that the intermediate solution of problem (30), which includes the cosines of the angles between the neutron and nucleus vectors (31) and (32), requires transformations of these cosines of angles to the "L"-system, i.e., several more relations transforming to the "L"-system the unit vectors specifying the direction of the neutron and nucleus velocity vectors after scattering.

**The law of inelastic neutron scattering taking into account thermal motions of the nuclei of the moderating medium**

As follows from the law of conservation of total energy for a closed system of a neutron and a nucleus stationary in L´ the system, the excitation energy of the composite nucleus is equal (Kamal, 2014; Marguet, 2018):

$$E^{*} = \varepsilon_n + \frac{A}{A+1} T_n^{(L')} \tag{33}$$

where: $\varepsilon_n$ - binding energy of the neutron in the composite nucleus; $T_n^{(L')}$ - - kinetic energy of the neutron before inelastic scattering $L'$ - system.

Note that fission reactions are not considered in this section. Here we consider the inelastic neutron scattering reaction on reproducing reactor fuel nuclides for which $\varepsilon_n$ less than the threshold fission energy of the nuclide $E_f$ such as uranium 238 ( $E_f^{U238}$ = 5.85 MeV and $\varepsilon_n^{U239}$ = 4.80 MeV (Rusov et al., 2015a; V. D. Rusov et al., 2011; Viktor Tarasov et al., 2023)[6-9]) and thorium 232 ( $E_f^{U238}$ = 5.90 MeV and $\varepsilon_n^{U239}$ = 4.79 MeV (Kamal, 2014; Marguet, 2018). Below in Section 4 are the fission reactions of the fissile reactor nuclides uranium 235 ( $E_f^{U235}$ = 5.75 MeV and = $\varepsilon_n^{U236}$ 6.55 (Rusov et al., 2015a; V. D. Rusov et al., 2011; Viktor Tarasov et al., 2023), uranium 233 ( $E_f^{U233}$ = 5.50 MeV and = $\varepsilon_n^{U234}$ 6.84 MeV (Kamal, 2014; Marguet, 2018)), and plutonium 239 ( $E_f^{Pu239}$ = 5.50 MeV and = $\varepsilon_n^{Pu240}$ 6.53 MeV (Kamal, 2014; Marguet, 2018)), for which $\varepsilon_n$ greater than the threshold fission energy of a nuclide $E_f$ , are taken into account by specifying the total number of neutrons generated by fission reactions per unit volume per unit time and the energy spectrum of neutron fissions.

Like any endothermic nuclear reaction, the inelastic neutron scattering reaction is a threshold nuclear reaction, i.e., inelastic scattering is possible not at any kinetic energy of the neutron, but at its kinetic energy exceeding the threshold $T_0^N$ This is confirmed by experiment, for example, according to, inelastic neutron scattering on heavy nuclei will be observed at neutron energies exceeding several hundred kiloelectronvolts, and on light nuclei - at energies exceeding one or several megaelectronvolts. This can also be seen in Fig. 3, which shows the neutron energy dependences of the neutron nuclear reaction microselections for uranium 238 at a temperature equal to 600 K. Note also that the threshold of inelastic neutron scattering reactions lies above the resonance range of neutron energies, which for reactor heavy nuclei is 1eV -100 keV and in which the discretization of the energy levels of nuclei is significantly manifested. This can also be seen in Fig. 3., since the dependence of the inelastic scattering cross section on the neutron energy has no resonances.

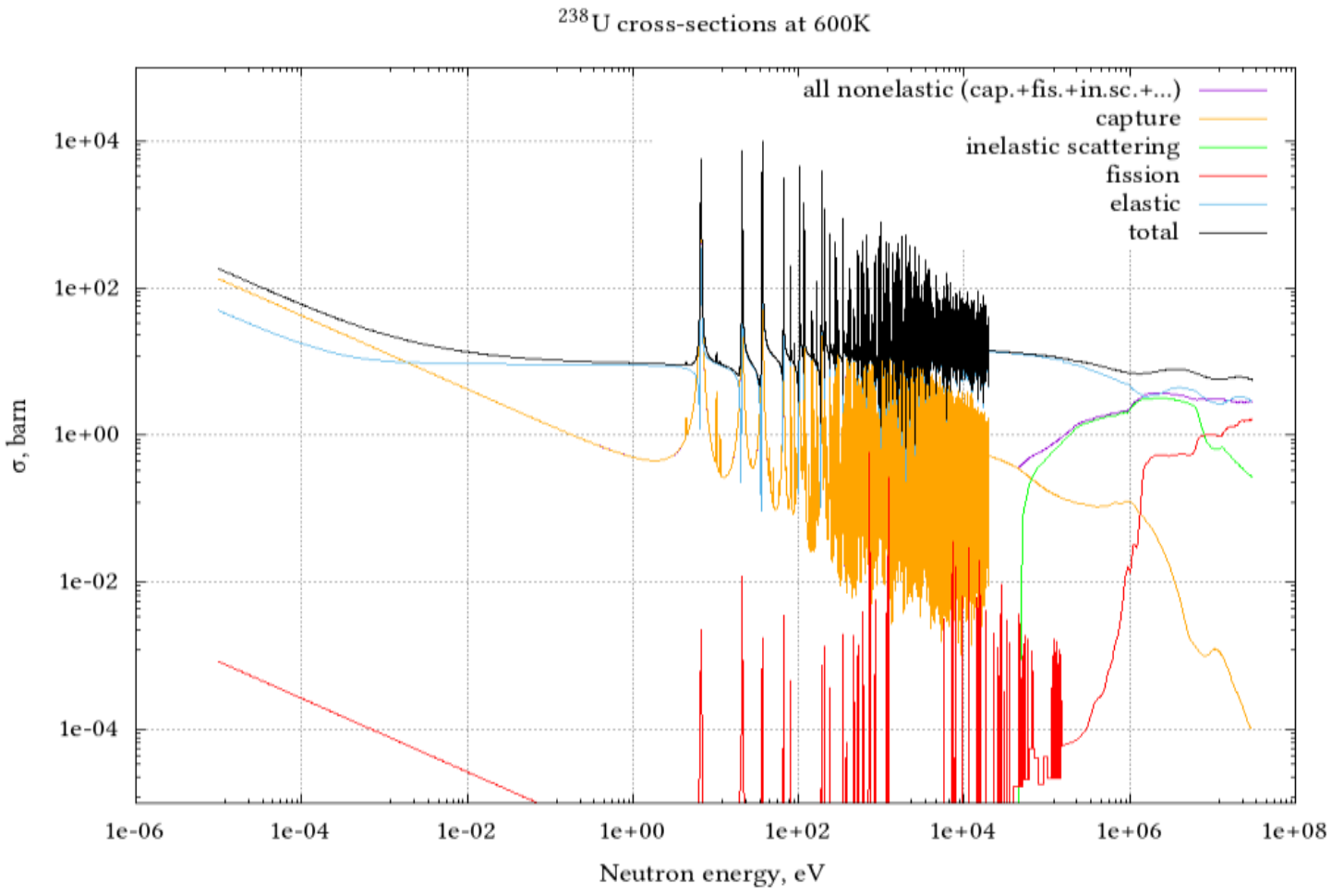

*Fig. 3.* Neutron energy dependences of micro cross sections of neutron nuclear reactions for uranium 238 at a temperature of 600 K.

Thus, since the inelastically scattered energy of the colliding neutrons changes from $T_0^N$ and up to $E_f - \varepsilon_n$ then the excitation energy of the compound nucleus due to the transfer of a part of the neutron kinetic energy can vary in the following interval:

$$\frac{A}{A+1} \cdot T_0^N \leq \frac{A}{A+1} \cdot \frac{1 \cdot \left(\overrightarrow{V_{10}}^{(L')}\right)^2}{2} < E_f - \varepsilon_n \tag{34}$$

At decay of a composite nucleus with formation of a moderating medium nucleus, in the unexcited state, a neutron, and of the γ -quantum (reaction scheme (2)), a part of the excitation energy of the compound nucleus $\varepsilon_n$ is spent for work against the strong interaction forces at neutron emission, and the remaining part of the excitation energy is equal to the sum of the kinetic energy of the system (nucleus and neutron) in the - system after scattering and the energy of the γ -quantum. Thus according to (33), (34), (21) and the above we obtain:

$$\frac{A}{A+1} \cdot T_n^{(L')} = \frac{A}{A+1} \frac{1 \cdot \left(\overrightarrow{V_{10}}^{(L')}\right)^2}{2} = \frac{\left(\overrightarrow{P_1}^{(Ts)}\right)^2}{2\mu'} + E_\gamma \tag{35}$$

For a stochastic set of inelastic neutron scattering reactions, since the kinetic energy of a neutron emitted by a decaying composite nucleus can vary randomly from zero to a maximum equal to (35) $\frac{A}{A+1} \frac{1 \cdot \left(\overrightarrow{V_{10}}^{(L')}\right)^2}{2}$ then correspondingly the energy of the emitted γ - quantum $E_\gamma$ varies in the interval:

$$0 \leq E_\gamma \leq \frac{A}{A+1} \frac{1 \cdot \left(\overrightarrow{V_{10}}^{(L')}\right)^2}{2} < E_f - \varepsilon_n \tag{36}$$

where $E_f$ is the threshold fission energy of the compound nucleus $(A+1, Z)$ (Rusov et al., 2012).

For uranium 238( $E_f^{U238}$ = 5.85 MeV and $\varepsilon_n^{U239}$ = 4.80 MeV (Kamal, 2014; Marguet, 2018)) and thorium 232 ( $E_f^{Th232}$ = 5.90 MeV and = $\varepsilon_n^{Th233}$ 4.79 MeV (Kamal, 2014; Marguet, 2018), the energy of the emitted γ - quantum $E_\gamma$ varies in the interval:

$$0 \leq E_\gamma^{U238} \leq \frac{A}{A+1} \frac{1 \cdot \left(\overrightarrow{V_{10}}^{(L')}\right)^2}{2} < 1{,}05 \text{ and } 0 \leq E_\gamma^{T238} \leq \frac{A}{A+1} \frac{1 \cdot \left(\overrightarrow{V_{10}}^{(L')}\right)^2}{2} < 1{,}11 \tag{37}$$

Uranium 238 and thorium 232 nuclei will fission when capturing neutrons with kinetic energies equal and greater than 1.05 MeV and 1.11 MeV, respectively.

The contribution to the neutron deceleration spectrum is made by a stochastic set of reactions of inelastic scattering of neutrons on the nuclei of the deceleration medium, with the kinetic energy of the γ -quantum (as well as the value of the neutron momentum modulus in the - system after scattering unambiguously related to it through relation (22)) varying randomly in the range of energies given by expression (36). There is no reason to consider that some value of kinetic energy of γ - quantum has a singular random character, i.e. that γ - quanta with such energy are somehow emitted more often than others. Therefore, we can assume that the random value of the energy of γ - quanta at inelastic scattering of a set of neutrons on the nuclei of the decelerating medium will be given by the equiprobability law, and the probability density is given according to (36) by the following expression:

$$\rho(E_\gamma) = \rho\left(\left|\overrightarrow{P_1}^{(Ts)}\right|\right) = \frac{1}{\left(\frac{A}{A+1}\right)\frac{\left(\overrightarrow{V_{10}}^{(L')}\right)^2}{2}} \tag{38}$$

For a sufficiently large stochastic population of neutrons and nuclei (for reactor physics the neutron flux density is more than 1013 n/cm²s the nuclei density ~1021 1/cm³) that have interacted through the inelastic scattering reaction, there will always be a stochastic sample of neutrons that have the same momentum vector before scattering $\overrightarrow{V_{10}}^{(L')}$ and, due to the isotropy of scattering, the same direction of the momentum vector. Let us denote it by $\vec{e}_1$ in the CD-system, in which the kinematics of inelastic scattering was considered above in Section 2. Then the mean value of the momentum modulus for neutrons inelastically scattered in the direction of $\vec{e}_1$ in the CD-system or equal to its average value of the velocity of such neutrons (since for the neutron A=1) under the equiprobability law is easily found as the sum of its maximum (at $E_f = 0$ and its minimum value (at $E_f = \frac{A}{A+1}T_n^{(L')}$ divided by two. Thus, using the expression for the momentum modulus (22) for the mean neutron momentum modulus and equal to it the mean square of the neutron velocity modulus $A_n = 1$ inelastically scattered in the direction $\vec{e}_1$ in the Ts-system, we get;

$$\overrightarrow{P_1}^{(Ts)} = 1 \cdot \overrightarrow{V_1}^{(Ts)} = \frac{1}{2}\left(\sqrt{2\mu'\left[\frac{\mu\left(V_{10}^{(L')}\right)^2}{2} + \frac{A}{A+1}\frac{\left(V_{10}^{(L')}\right)^2}{2}\right]} + \sqrt{2\mu'\left[\frac{\mu\left(V_{10}^{(L')}\right)^2}{2}\right]}\right) =$$

$$= \frac{1}{2}\left(2\mu'\frac{3\mu\left(V_{10}^{(L')}\right)^2}{2}\right) \tag{39}$$

Since according to the estimates (37) the maximum energy of γ - quantum $E_\gamma$ ~ 1 MeV, 1 a.e.m. ≈ 931.50 MeV, then $\mu' = \frac{\left(A+{}^{E_\gamma}/_{c^2}\right)\cdot 1}{\left(A+{}^{E_\gamma}/_{c^2}\right)+1} \approx \mu = \frac{A\cdot 1}{A+1}$ and then from (31) we obtain:

$$\overrightarrow{P_1}^{(Ts)} = \overrightarrow{V_1}^{(Ts)} = \frac{1}{2}\left(\sqrt{\frac{4\mu\left(V_{10}^{(L')}\right)^2}{2}} + \sqrt{\frac{2\mu^2\left(V_{10}^{(L')}\right)^2}{2}}\right) = \frac{(\sqrt{2}+1)\mu(V_{10}^{(L')})}{2} \text{ and}$$

$$\left(\overrightarrow{V_1}^{(Ts)}\right)^2 \approx \frac{\left(\sqrt{2}+1\right)^2\mu\left(V_{10}^{(L')}\right)}{4} = \frac{\left(\sqrt{2}+1\right)^2 A^2\left(V_{10}^{(L')}\right)^2}{4(A+1)^2} = \frac{A^2B^2\left(V_{10}^{(L')}\right)^2}{(A+1)^2}$$

where: $B^2 = \frac{\left(\sqrt{2}+1\right)^2}{4}$ (40)

Knowing the mean value of the neutron velocity modulus after scattering in the CD-system, we can, similarly to expression (23) from Section 2, multiply it by the unit vector of neutron escape in the $\vec{e}_1$ CD-system to write an expression for the mean velocity vector of neutrons escaping in the direction $\vec{e}_1$ after inelastic scattering:

$$\overrightarrow{V_1}^{(Ts)} = \frac{AB}{(A+1)} V_{10}^{(L')} \cdot \overrightarrow{e_1} \quad (41)$$

Squaring the expression (41) for the value of the square of the mean velocity of neutrons flown in the direction $\overrightarrow{e_1}$ after scattering in the L'-system, we obtain the following expression:

$$\left(V_{10}^{(L')}\right)^2 = \left(V_{10}^{(Ts)}\right)^2 + (V_{10}^{(L')} \cdot \frac{1}{A+1})^2 + \frac{2\cdot(V_1^{(Ts)})\cdot(V_{10}^{(L')}}{A+1} \cdot \cos\theta \quad (42)$$

where, as well as above in section $2\theta$ , is the neutron departure angle in the system of the center of inertia "C" (Fig. 2), counted from the direction of the neutron velocity vector before the interaction in the "L" system. Indeed, according to (15), the direction of the vector $V_{Ts}^{(L')}$ (Fig. 2) coincides with the direction of the vector .

Substituting in (33) the expression for $V_{Ts}^{(L')}$ (41) we obtain the following expression:

$$\left(V_{Ts}^{(L')}\right)^2 = \frac{\left(V_{10}^{(L')}\right)^2 \cdot (A^2B^2 + 2AB\cos\theta + 1)^2}{(A+1)^2} \quad (43)$$

Further, due to the same mathematical structure of expression (43) and expression (26) from Section 2, we can make calculations similar to the calculations from Section 2 after formula 26.

From the relation (43) we can find the ratio of the square of the average neutron velocity after inelastic scattering $\bar{E}_2^{(L')}$ and the square of the neutron velocity before scattering$E_2^{(L')}$ (inelastic scattering of a set of neutrons having the same velocity model before scattering is considered) in the laboratory coordinate system "L'", which is also equal to the ratio of the average neutron kinetic energy after scattering and the neutron kinetic energy before scattering:

$$\frac{\left(\overrightarrow{V_1}^{(L)}\right)^2}{\left(\overrightarrow{V_{10}}^{(L)}\right)^2} = \frac{E_2^{(L')}}{E_1^{(L')}} = \frac{A^2B^2 + 2AB\cos\theta + 1}{(A+1)^2} \quad (44)$$

If we introduce the inelastic scattering coefficients as follows:

$$\tilde{\alpha}_1^2 = (\frac{AB-1}{AB+1})^2 \text{ and} \tilde{\alpha}_1^2 = (\frac{AB+1}{AB+1})^2 \quad (45)$$

then expression (44) can be transformed to the following form:

$$\frac{E_2^{(L')}}{E_1^{(L')}} = \frac{\tilde{\alpha}_1^2}{2}[(1+\tilde{\alpha}_1^2) + (1-\tilde{\alpha}_1^2)\cos\theta] \quad (46)$$

Now, expressing the vector of the average neutron velocity after scattering in the CD-system through the analogous vector of the average neutron velocity after scattering in the L-system, we can write expression (46) as the following expression:

$$\frac{\left(\overrightarrow{V_1}^{(L)} - \overrightarrow{V_{20}}^{(L)}\right)^2}{\left(\overrightarrow{V_{10}}^{(L)} - \overrightarrow{V_{20}}^{(L)}\right)^2} = \frac{E_2^{(L')}}{E_1^{(L')}} = \frac{\tilde{\alpha}_1^2}{2}[(1+\tilde{\alpha}_1^2) + (1-\tilde{\alpha}_1^2)\cos\theta] \quad (47)$$

From expression (47) for inelastic scattering we obtain an expression similar to the expression for elastic scattering obtained in (Tarasov, 2017):

$$\frac{\left(\overrightarrow{V_1}^{(L)}\right)^2}{\left(\overrightarrow{V_{10}}^{(L)}\right)^2} = \frac{E_1^{(L')}}{E_{10}^{(L')}} = \frac{\tilde{\alpha}_1^2}{2}[(1+\tilde{\alpha}_1^2) + (1-\tilde{\alpha}_1^2)\cos\theta] \cdot \left[1 - 2\frac{\overrightarrow{V_1}^{(L)}\overrightarrow{V_{20}}^{(L)}\cos\beta}{\left(\overrightarrow{V_{10}}^{(L)}\right)^2} + \frac{\left(\overrightarrow{V_{20}}^{(L)}\right)^2}{\left(\overrightarrow{V_{10}}^{(L)}\right)^2}\right] +$$

$$2\frac{\overrightarrow{V_1}^{(L)},\overrightarrow{V_{20}}^{(L)}\cos\beta}{\left(\overrightarrow{V_{10}}^{(L)}\right)^2} + \frac{\left(\overrightarrow{V_{20}}^{(L)}\right)^2}{\left(\overrightarrow{V_{10}}^{(L)}\right)^2} = \frac{\tilde{\alpha}_1^2}{2}[(1+\tilde{\alpha}_1^2) + (1-\tilde{\alpha}_1^2)\cos\theta] - \frac{\tilde{\alpha}_1^2}{2}[(1+\tilde{\alpha}_1^2) + (1-$$

$$\tilde{\alpha}_1^2)\cos\theta]\frac{\overrightarrow{V_{10}}^{(L)}\overrightarrow{V_{20}}^{(L)}\cos\beta}{\left(\overrightarrow{V_{10}}^{(L)}\right)^2} + \frac{\overrightarrow{V_1}^{(L)}\overrightarrow{V_{20}}^{(L)}\cos\gamma}{\left(\overrightarrow{V_{10}}^{(L)}\right)^2} - \left\{1 - \frac{\tilde{\alpha}_1^2}{2}[(1+\tilde{\alpha}_1^2) + (1-\tilde{\alpha}_1^2)\cos\theta]\right\}\frac{1\cdot E_{20}^{(L)}}{A\cdot E_{10}^{(L)}} \quad (48)$$

where: $\cos\beta$ is the cosine of the angle between the vectors $\overrightarrow{V_{10}}^{(L)}$ and $\overrightarrow{V_{20}}^{(L)}$, which is expressed through the scalar product ($\overrightarrow{V_{10}}^{(L)}$, $\overrightarrow{V_{20}}^{(L)}$) of these vectors as follows:

$$\cos\beta = \frac{\left(\overrightarrow{V_{10}}^{(L)}, \overrightarrow{V_{20}}^{(L)}\right)}{\left|\overrightarrow{V_{10}}^{(L)}\right|\left|\overrightarrow{V_{20}}^{(L)}\right|} = \frac{\left(\overrightarrow{V_{10}}^{(L)}, \overrightarrow{V_{20}}^{(L)}\right)}{\overrightarrow{V_{10}}^{(L)}\overrightarrow{V_{20}}^{(L)}} \tag{49}$$

where: $\cos\gamma$ is the cosine of the angle between the vectors $\overrightarrow{V_1}^{(L)}$ and $\overrightarrow{V_{20}}^{(L)}$:

$$\cos\beta = \frac{\left(\overrightarrow{V_1}^{(L)}, \overrightarrow{V_{20}}^{(L)}\right)}{\left|\overrightarrow{V_1}^{(L)}\right|\left|\overrightarrow{V_{20}}^{(L)}\right|} = \frac{\left(\overrightarrow{V_1}^{(L)}, \overrightarrow{V_{20}}^{(L)}\right)}{\overrightarrow{V_1}^{(L)}\overrightarrow{V_{20}}^{(L)}} \tag{50}$$

Since according to expression (48), the average kinetic energy of the considered neutron sample after scattering is $\bar{E}_1^{(L)}$ is a function of several independent variables $\bar{E}_1^{(L)} = \int(\theta, \beta, \gamma, E_\gamma^L)$ and in the considered case of scattering of neutrons emitted by an isotropic source on chaotically moving nuclei of the medium, the probability density function for a set of random variables $P\theta, \beta, \gamma, E_\gamma^L$ by virtue of their independence is equal to $P\left(\theta, \beta, \gamma, E_\gamma^L\right) = P(\theta) \cdot P(\beta) \cdot P(\gamma) \cdot P(E_\gamma^L)$ then the probability that the considered sample of neutrons having kinetic energy $E_{10}^{(L)}$ before scattering on nuclei in the laboratory coordinate system “L”, after scattering will have average kinetic energy in the range from $E_1^{(L)}$ before $\bar{E}_1^{(L)} + d\bar{E}_1^{(L)}$ can be written in the following form:

$$P\left(\bar{E}_1^{(L)}\right) dE_1^{(L)} = \left[\iiint_{\theta,\beta,\gamma,E_\gamma^L} \int \delta\left(\bar{E}_1^{(L)} - \int(\theta, \beta, \gamma, E_\gamma^L)\right)(\theta) \cdot P(\beta) \cdot P(\gamma) \cdot P(E_\gamma^L) d\theta d\beta d\gamma dE_\gamma^L\right] d\bar{E}_1^{(L)} \tag{51}$$

However, it is possible to proceed further in another way and expression (51) will not be used hereafter.

Since the decay of a composite nucleus, as is known from nuclear physics, does not depend on the prehistory of its formation, inelastic neutron scattering in the coordinate system of the center of inertia is spherically symmetric (isotropic).

Then for $P(\theta)d\theta$ get:

$$P(\theta)d\theta = \int_0^{2\pi} [P(\theta, \varphi)d\theta]\, d\varphi = \int_0^{2\pi} \frac{r \sin\theta d\theta \cdot r d\theta}{4\pi r^2} = \frac{\sin\theta d\theta \cdot r d\theta}{4\pi} \int_0^{2\pi} d\varphi = \frac{1}{2}\sin\theta d\theta \tag{52}$$

where through $\varphi$ denotes the azimuthal angle of ordinary spherical coordinates $r, \theta, \varphi$ introduced in the coordinate system of the center of inertia.

Since the thermal motion of the moderating medium nuclei is chaotic and the neutron source is isotropic (the neutron source emits a group of neutrons with a given energy with an isotropic spatial distribution of the directions of their velocity vectors), the distribution of velocity vector directions in space for neutrons after inelastic scattering is given by the equiprobability law of distribution by angles $\beta$ and $\gamma$ included in expression (48), i.e., it is also spherically symmetric (isotropic), then we obtain similarly to the previous one:

$$P(\beta)d\beta = \frac{1}{2}\sin\beta d\beta \tag{53}$$

$$P(\beta)d\beta = \frac{1}{2}\sin\beta d\beta \tag{54}$$

Let us average the average kinetic energy of the considered sample of neutrons after inelastic scattering on the nucleus, given by expression (48), over the spherically symmetric distribution of velocity directions of thermal motion of nuclei, the moderating medium, and over an isotropic neutron source (over the isotropic spatial distribution of velocity vectors of neutrons with a given energy emitted by the neutron source) and obtain the following expression:

$$\bar{\bar{E}}_1^{(L)} = \iint_0^{\pi} \bar{E}_1^{(L)} P(\beta) d\beta P(\gamma) d\gamma = \bar{E}_{20}^{(L)} \left\{ \frac{\tilde{\alpha}_1^2}{2} [(1+\tilde{\alpha}_1^2) + (1-\tilde{\alpha}_1^2)\cos\theta] - \left[1 - \frac{\tilde{\alpha}_1^2}{2}[(1+\tilde{\alpha}_1^2) + (1-\tilde{\alpha}_1^2)\cos\theta]\right] \frac{1 \cdot E_{20}^{(L)}}{A \cdot E_{10}^{(L)}} \right\} \quad (55)$$

In this expression $\bar{E}_{10}^{(L)}$ is the neutron energy averaged along the neutron momentum directions for an isotropic neutron source, which coincides with the neutron energy$E_{10}^{(L)}$, emitted by the source, i.e.,$\bar{E}_{10}^{(L)} = E_{10}^{(L)}$and $E_y^{(L)}$is given by the Maxwell distribution for the nuclei of the moderating medium, which depends on the temperature of the moderating medium as a parameter and has the following form:

$$P\left(E_y^{(L)}\right) \mathrm{d}E_y^{(L)} = \frac{2}{\sqrt{\pi (kT)^3}} e^{-\frac{E_y^{(L)}}{kT}} \sqrt{E_y^{(L)}} dE_y^{(L)} \quad (56)$$

Let us average expression (55) over the Maxwell distribution of thermal motion of the decelerating medium nuclei (56), taking into account that$\bar{E}_{10}^{(L)} = E_{10}^{(L)}$ and if we use the known result$E_y^{(L)} = \int_0^{\infty} E_y^{(L)} P_M\left(E_y^{(L)}\right) dE_y^{(L)} = \frac{3}{2} kT$ then we obtain the following expression:

$$\bar{\bar{E}}_1^{(L)} = \bar{E}_{10}^{(L)} \left\{ \frac{\tilde{\alpha}_2^2}{2} [(1+\tilde{\alpha}_1^2) + (1-\tilde{\alpha}_1^2)\cos\theta] - \left[1 - \frac{\tilde{\alpha}_2^2}{2}[(1+\tilde{\alpha}_1^2) + (1-\tilde{\alpha}_1^2)\cos\theta]\right] \frac{\frac{3}{2}kT}{A \cdot E_{10}^{(L)}} \right\} \quad (56.1)$$

Thus, since the functional relationship between $\bar{\bar{E}}_1^{(L)}$and $\theta$ is unambiguous as follows from relation (56), then the probability that the neutrons having kinetic energy $E_{10}^{(L)}$ before scattering on a nucleus in the laboratory coordinate system "L", after scattering on chaotically moving nuclei of the decelerating medium will have kinetic energy averaged over the thermal motion of the nuclei, $\bar{\bar{E}}_1^{(L)}$ between $\bar{\bar{E}}_1^{(L)}$ before $\bar{\bar{E}}_1^{(L)} + d\bar{\bar{E}}_1^{(L)}{}^{P(\bar{\bar{E}}_1^{(L)})d\bar{\bar{E}}_1^{(L)}}$ is given by the distribution $P(\theta)d\theta$ (2) and therefore we obtain the following relation (here for simplicity of writing we omit signs of averaging and laboratory coordinate system "L", i.e., we denote by $P(\bar{\bar{E}}_1^{(L)})dE_{10}^{(L)} = P(E_1)dE_1$

$$P(E_1)dE_1 = P(\theta)d\theta = P(\theta)\left|\frac{d\theta}{dE_1}\right| dE_1 = \frac{1}{2}\sin\theta \left| \frac{1}{E_{10}^{(L)}\left[\frac{\tilde{\alpha}_2^2}{2}\left[(1+\tilde{\alpha}_1)\sin\theta + (1-\alpha_1)\sin\theta \frac{\frac{3}{2}kT}{A \cdot E_{10}^{(L)}}\right]\right]} \right| = \frac{dE_1}{\left[E_{10}^{(L)} + \frac{1}{A}\frac{3}{2}kT\right]\tilde{\alpha}_2^2(1-\alpha_1)} \quad (57)$$

Thus, the law of inelastic scattering of neutrons, taking into account the thermal motion of the nuclei of the moderating medium, is obtained:

$$P(E_1)dE_1 = \frac{dE_1}{\left[E_{10}^{(L)} + \frac{1}{A}\frac{3}{2}kT\right]\tilde{\alpha}_2^2(1-\alpha_1)} \text{if} \tilde{\alpha}_2^2 \alpha_1 (E_{10}^{(L)} + \frac{1}{A} \cdot \frac{3}{2} kT) \le E_1 \le (E_{10}^{(L)} + \frac{1}{A} \cdot \frac{3}{2} kT)$$

$$P(E_1) = 0 \text{if } E_1 < \tilde{\alpha}_2^2 \alpha_1 (E_{10}^{(L)} + \frac{1}{A} \cdot \frac{3}{2} kT) \text{ and } E_1 > (E_{10}^{(L)} + \frac{1}{A} \cdot \frac{3}{2} kT) \quad (58)$$

In conclusion, we emphasize that the inelastic scattering law (58) is written for the averaged neutron energy after scattering $E_1$ Neutron energy is averaged according to the thermal (chaotic) motion of the nuclei of the retarding medium and the isotropy of the neutron source. And as follows from the law of scattering (58), all neutrons emitted by an isotropic source of neutrons and having the energy to scatter on the nuclei of the retarding medium $E_{10}^{(L)}$ with the probability given by expression (58), after inelastic scattering on the nuclei of the decelerating medium will have energy averaged over the thermal (chaotic) motion of the nuclei of the decelerating medium and the isotropy of the neutron source $E_1$.

As we will see below, the obtained scattering law in the form (58) allows us to obtain expressions for the neutron flux density and energy spectra of neutrons decelerating in different decelerating media, taking into account the temperature of the decelerating medium.

**Neutron deceleration in decelerating neutron-absorbing media containing nuclei of several varieties**

According to (Tarasov, 2017) it is possible to find an analytical solution to the stationary balance equation for slowing down neutrons, which is the neutron flux density. In the case considered in (Tarasov, 2017), the expression for the flux density of slowing elastically scattered neutrons has the following form:

$$\Phi_1(E)=\frac{\int_E^\infty \frac{\Sigma_{el}(E')}{\Sigma_t(E')}Q(E')\cdot\left\{\exp\left[\int_{E'}^\infty -\frac{\Sigma_a(E'')dE''}{|\bar{\xi}_{el}|\Sigma_t(E'')\left[E''+\frac{1}{A}\frac{3}{2}kT\right]}\right]\right\}dE'}{\left[E+\frac{1}{A}\frac{3}{2}kT\right]\Sigma_t(E)|\bar{\xi}_{el}|}+\frac{\frac{\Sigma_{el}(E)}{\Sigma_t(E)}Q(E)}{\Sigma_t(E)} \tag{59}$$

where: $Q(E)$ - number of generated neutrons with energy $(A,Z)$ per unit volume per unit time,

$\Sigma_{el}$ - full macroscattering elastic scattering of the decelerating fissile medium ( $\Sigma_{el}=\sum_i \Sigma_{el}^i$ where $\Sigma_{el}^i$ - macrosegmentation of the nuclide included in the decelerating medium), $\Sigma_t=\Sigma_s+\Sigma_a$ full macro section

$\Sigma_s$ - the full macroscattering cross section of the fissile medium,

$\Sigma_a$ - the full macro section of neutron absorption,

$|\bar{\xi}_{el}|$ - - modulus of the mean logarithmic energy decrement for elastic scattering ${}_0^1n$ - which is defined similarly to the standard theory of neutron slowing down [1, 20], but through the new law of elastic neutron scattering from [20].

Recall that $\Sigma_s=\Sigma_{el}+\Sigma_{in}$ where $\Sigma_{el}=\Sigma_p+\Sigma_{rs}$ $\Sigma_{el}=\Sigma_p+\Sigma_{rs}$ – - full macro section of elastic scattering $\Sigma_p$ -- full macro section of potential scattering $\Sigma_{rs}$ -- full macro section of resonant scattering $\Sigma_{in}$ -- macro section of inelastic scattering (Rusov et al., 2012).

Included in (59) is an expression for the resonant non-absorption probability function for the neutron (Marguet, 2018; Tarasov, 2017), which now also contains the temperature of the moderating medium:

$$\varphi(E')=\exp\left[-\int_E^\infty \frac{\Sigma_a(E')dE'}{|\bar{\xi}_{el}|\Sigma_t(E')\left[E'+\frac{1}{A}\cdot\frac{3}{2}kT\right]}\right] \tag{60}$$

Knowing the inelastic scattering law (58) and making calculations similar to those made for the elastic scattering law in (Tarasov, 2017), we obtain an expression for the flux density of slowing down inelastically scattered neutrons in the following form:

$$\Phi_2(E)=\frac{\int_E^\infty \frac{\Sigma_{in}(E')}{\Sigma_t(E')}Q(E')\cdot\left\{\exp\left[\int_{E'}^\infty -\frac{\Sigma_a(E'')dE''}{|\bar{\xi}_{in}|\Sigma_t(E'')\left[E''+\frac{1}{A}\cdot\frac{3}{2}kT\right]}\right]\right\}dE'}{\left[E+\frac{1}{A}\cdot\frac{3}{2}kT\right]\Sigma_t(E)|\bar{\xi}_{in}|}+\frac{\frac{\Sigma_{in}(E)}{\Sigma_t(E)}Q(E)}{\Sigma_t(E)} \tag{61}$$

where the mean logarithmic energy decrement for inelastic scattering is $\xi_{in}$ introduced analogously to the standard theory of neutron slowing down (Kamal, 2014; Tarasov, 2017), but already through the law of inelastic neutron scattering (58) and having made calculations analogous to those performed for $\xi_{el}$ and given in (Tarasov, 2017)[20], we obtain the following expression:

$$\xi_{in} = \frac{\tilde{\alpha}_1}{(1-\tilde{\alpha}_1)} \ln\left(\tilde{\alpha}_2^2 \tilde{\alpha}_1\right) + \frac{\left(1-\tilde{\alpha}_2^2 \tilde{\alpha}_1\right)}{\tilde{\alpha}_2^2 (1-\tilde{\alpha}_1)} \left[1 + \ln\left(\frac{E}{E + \frac{1}{A} \cdot \frac{3}{2} kT}\right)\right] \quad (62)$$

Thus, taking into account the contribution of elastic and inelastic neutron scattering, we can find an expression for the total flux density of decelerating neutrons:

$$\Phi(E) = \Phi_1(E) + \Phi_2(E) \quad (63)$$

And the neutron spectrum is given by the standard expression:

$$\rho(E)dE = \frac{n(E)}{\int_0^\infty n(E)dE} = \frac{\frac{1}{\sqrt{2E}}\Phi(E)dE}{\int_0^\infty \frac{1}{\sqrt{2E}}\Phi(E)dE} \quad (64)$$

For reactor fissile media $Q(E)$ - is given by the fission spectrum of the fissile nuclide (or their combination), which according to [2, 20,25] can be given by the following expression:

$$Q(E) = \tilde{Q} \cdot c \exp(-aE) sh\sqrt{bE} \quad (65)$$

where c, a and b are constants, which are presented below in Table 1, E is the neutron energy unmeasured at 1 MeV $\tilde{Q} = \int_0^\infty Q(E)dE$ - the total number of neutrons generated neutrons per unit volume per unit time.

The Figure 4 shows for the energy spectrum of neutrons decelerating in a hydrogen medium obtained by computer calculation using expression (63). In the calculation, the source of fission neutrons was given by expression (65) for uranium 235 (Table 1), and the temperature of the moderating medium was 1200 K. Elastic neutron scattering microsections on hydrogen were taken from the ENDF/B-VII.0 database (Chadwick et al., 2011).

The graph (Fig. 5) shows the neutron energy spectrum (0 - 5 MeV) calculated using the GEANT code (Agostinelli et al., 2003), which implements the Monte Carlo method. In the calculation, the source of fission neutrons was given by expression (65) for uranium 235, with the temperature of the hydrogen moderator equal to 1000K.

So, a comparative analysis of the neutron spectra for the hydrogen moderator presented in (Fig. 4) and (Fig. 5) demonstrates their good agreement.

**Table 1.** Constant setting fission spectra for the main reactor fissile nuclides [24].

| Constant | $U^{235}$ | $Pu^{239}$ | $U^{233}$ | $Pu^{241}$ |
|---|---|---|---|---|
| a | 1,036 | 1 | 1,05±0,03 | 1,0±0,05 |
| b | 2,29 | 2 | 2,3±0,10 | 2,2±0,05 |
| c | 0,4527 | $\sqrt{\frac{2}{\pi e}}$ | 0,46534 | 0,43892 |

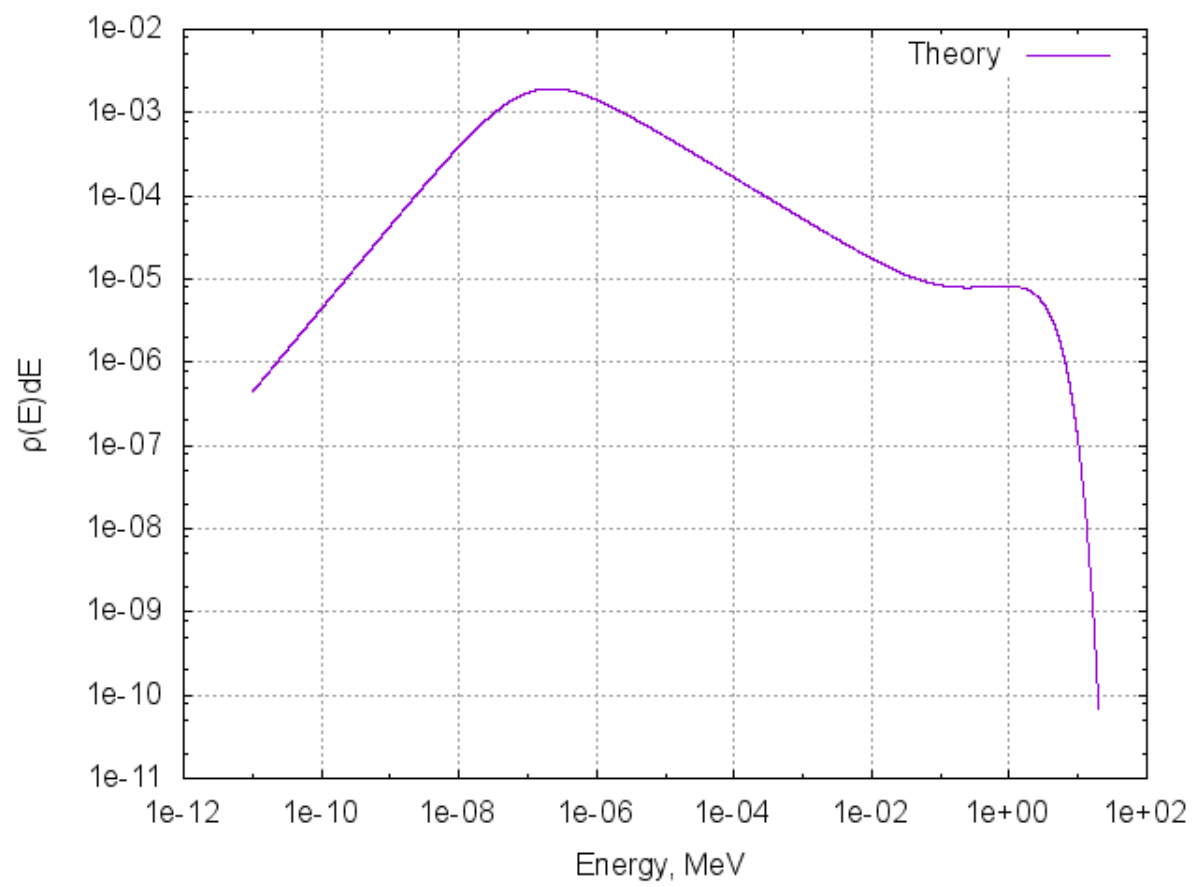

*Fig. 4.* Neutron energy spectra calculated from expression (63) and the fission neutron source given by expression (65) for uranium 235, hydrogen moderating medium temperature 1200 K

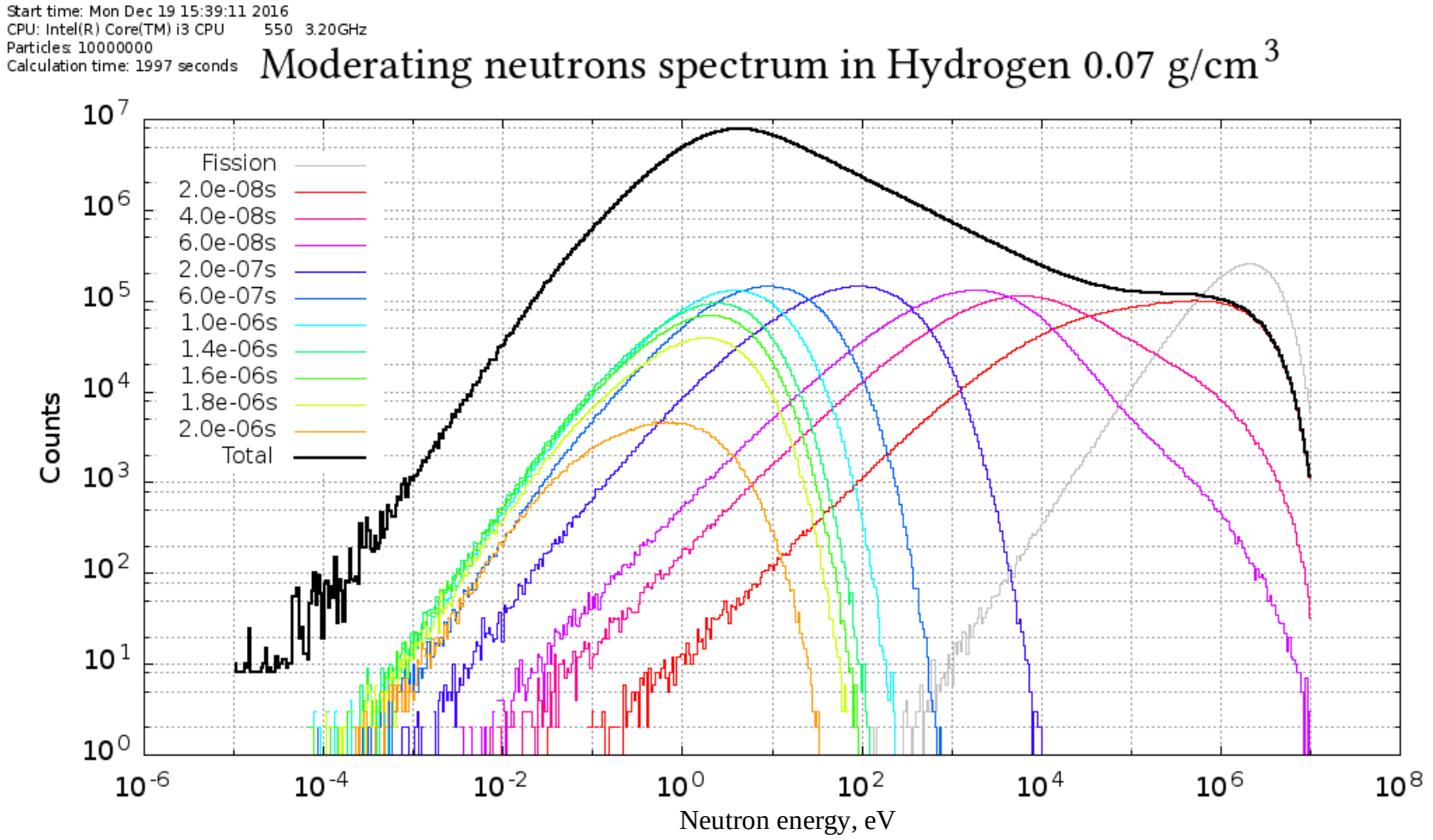

*Fig. 5.* Neutron energy spectrum calculated with the GEANT4 code for the fission neutron source given by expression (65) for uranium 235, with the temperature of the hydrogen moderator equal to 1000K

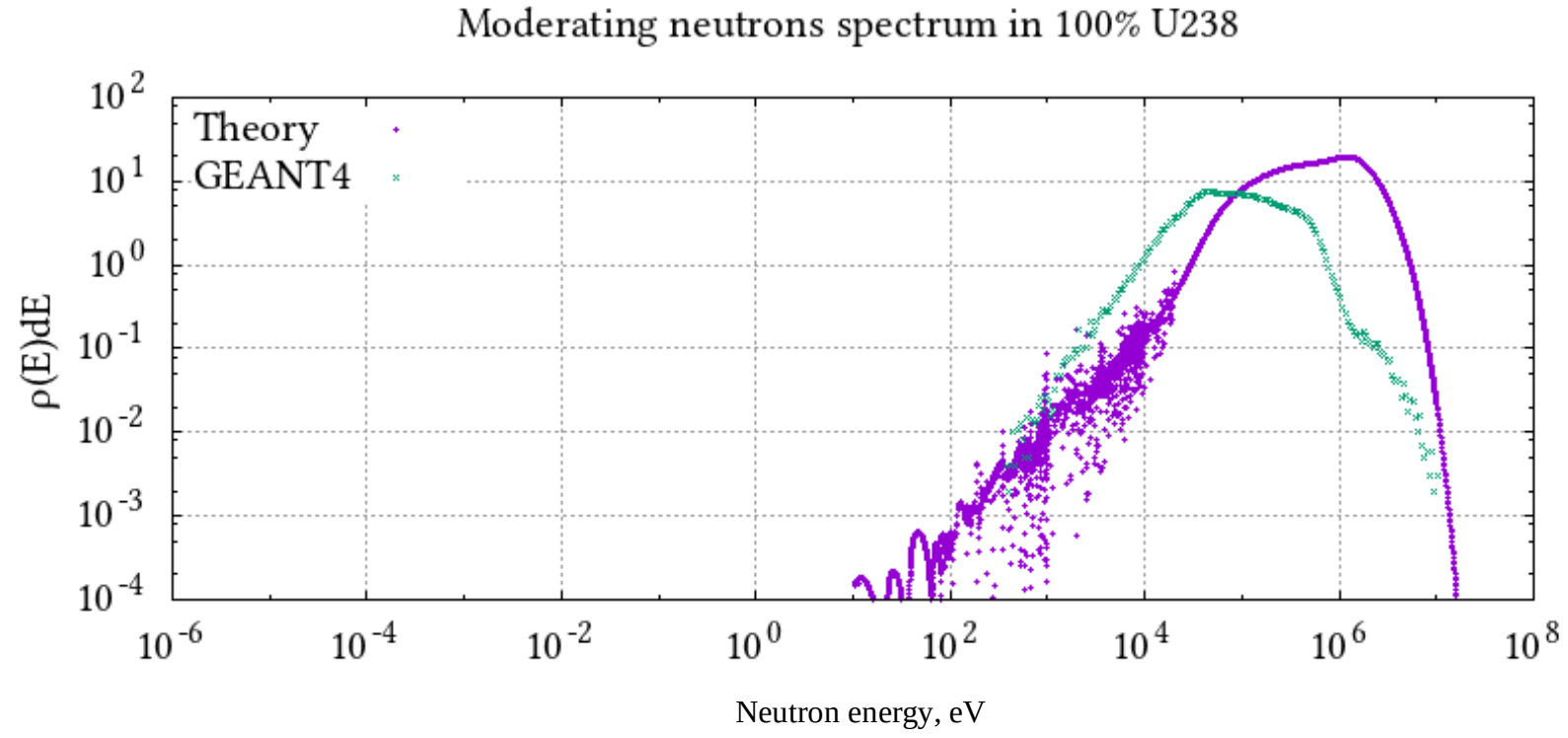

*Fig. 6.* Neutron energy spectrum for the uranium-carbon medium (100% uranium 238) calculated by expression (63) and using the GEANT4 code with the fission neutron source given by expression (65) at the moderator temperature equal to 600 K.

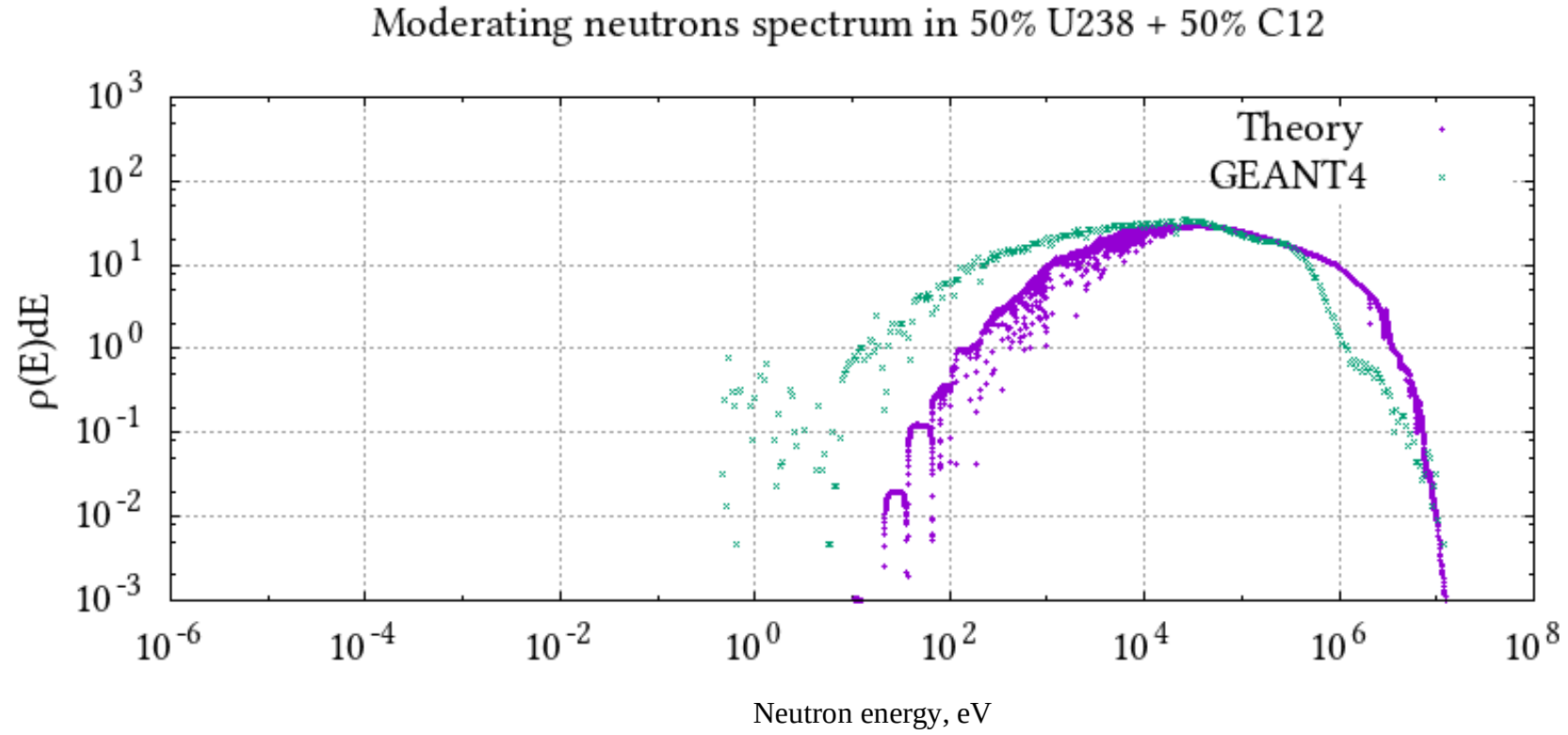

*Fig. 7.* Neutron energy spectrum for the uranium-carbon medium (50% uranium 238 and 50% carbon C12) calculated using expression (63) and the GEANT4 code with the fission neutron source given by expression (65) at the moderator temperature equal to 600 K.

We emphasize, as it was already mentioned in our previous paper (Tarasov, 2017), the analytical expression for the decelerating neutron spectrum obtained in the framework of the new theory of neutron slowing describes the full neutron spectrum, i.e, in the whole range of possible neutron energies and, depending on the composition of the decelerating medium and its temperature, may have either two maxima (high-energy and low-energy) or one (either high-energy - fast neutrons, or low-energy - thermal neutrons, or in the intermediate energy region - intermediate neutrons). This transformation of the theoretical spectrum is well demonstrated by the following calculated plots for the spectra of neutrons decelerating in uranium-carbon media having different uranium and carbon compositions.

The calculations were performed for different compositions of the uranium-carbon medium with different model percentages of uranium 238 and carbon 12, at a temperature of 600 K. So Fig. 6 shows neutron energy spectrums for the uranium medium (100% uranium 238) calculated by theoretical expression (63) and using the GEANT4 code with the fission neutron source given by expression (65) at the moderator temperature equal to 600 K. Fig. 7 shows neutron energy spectrums for the uranium-carbon medium (50% uranium 238 and 50% carbon C12) calculated using theoretical expression (63) and the GEANT4 code with the fission neutron source given by expression (65) at the moderator temperature equal to 600 K. Fig. 8 shows neutron energy spectrums for the uranium-carbon medium (20% uranium 238 and 80% carbon C12) calculated using theoretical expression (63) and the GEANT4 code with the fission neutron source given by expression (65) at the moderator temperature equal to 600 K. Fig. 9 shows neutron energy spectrums for the uranium-carbon medium (10% uranium 238 and 90% carbon C12) calculated using expression (63) and the GEANT4 code with the fission neutron source given by theoretical expression (65) at the moderator temperature equal to 600 K. Fig. 10 shows neutron energy spectrums for the uranium-carbon medium (5% uranium 238 and 95% carbon C12) calculated using expression (63) and the GEANT4 code with the fission neutron source given by expression (65) at the moderator temperature equal to 600 K. Fig. 11 shows neutron energy spectrums for the uranium-carbon medium (1% uranium 238 and 99% carbon C12) calculated by theoretical expression (63) and using the GEANT4 code with the fission neutron source given by expression (65) at the moderator temperature equal to 600 K. Fig. 12 shows neutron energy spectrums for a carbon medium (100% carbon C12) calculated by theoretical expression (63) and using the GEANT4 code with the fission neutron source given by expression (65) at the moderator temperature equal to 600 K.

A comparative analysis of the neutron spectra for uranium-carbon moderator media, presented in Figs. 6–12, demonstrates good agreement between the spectra calculated according to the developed moderation theory and the spectra calculated by the Monte Carlo method.

The presented spectra clearly demonstrate the transformation of the neutron spectra from the fast to the thermal spectrum depending on the increase in the amount of carbon in the uranium-carbon moderating mixture.

Particular attention should be paid to the presence of a sharp narrow maximum (ejection) in the low-temperature part of the theoretical neutron moderation spectrum, which is clearly visible in Figs. 9–12. It is caused by the transition of the mean logarithmic energy decrement $\xi$ through zero, since at $\xi=0$ the neutron flux density ((59) or (61)), tends to infinity. Indeed, as was shown in our article (Tarasov, 2017) for the new scattering law, the decrement does not remain constant, as in the standard moderation theory, but depends on the neutron energy and the temperature of the medium (for example, (62)): at neutron energies lower than thermal energies, the energy decrement $\xi$ passes through zero and becomes negative and as the neutron energy approaches zero it tends to minus infinity. That the mean logarithmic decrement of energy $\xi$ becomes negative, indicates that neutrons, when interacting with nuclei of the medium, possessing kinetic energy of thermal motion, with a high probability do not lose their energy (slowing down), but receive it. Thus, neutrons, when interacting with nuclei of the moderating medium, are, as it were, thrown off from their zero energy and the new theory of slowing down naturally includes a description of the process of neutron thermalization.

Let us consider the point of the spectrum corresponding to the mean logarithmic energy decrement equal to zero, i.e., the point at which the theoretical expression for the neutron flux diverges. According to the definition of the mean logarithmic energy decrement (Kamal, 2014; Marguet, 2018; Tarasov, 2017), in this case the energy of the neutrons before and after the interaction remains unchanged, i.e., the neutrons that received this energy during the moderation process or emitted by the neutron source with this energy, on average, do not change it during the further process of interaction with the nuclei of the moderating medium. This should lead to the accumulation of neutrons with such energy. However, if we consider the kinematics of two-particle elastic scattering, then the case where the lighter incident particle after scattering would have the same energy as before scattering corresponds on the momentum diagram to only one of the two points usually related to the case of a head-on collision, determined by the condition: both particles after the collision move along the same straight line in the L-system, that is $\bar{P}_1^{(L)} || \bar{P}_2^{(L)}$. Where $\bar{P}_1^{(L)}$and $\bar{P}_2^{(L)}$- the momenta of the incident and scattering particles of the L-system after interaction. This point corresponds to the solution $\bar{P}_1^{(L)} = \bar{P}_{10}^{(L)}$ and $\bar{P}_2^{(L)} = 0$.

Where $\bar{P}_{10}^{(L)}$-- the momentum of the incident particle to the L-system after the interaction. Usually this solution is considered as possible, but it is completely unphysical. Indeed, the scattering of one classical particle on another is considered, and the momentum vector does not change. This simply cannot be. Thus, for elastic processes, the neutron energy at which the energy decrement vanishes should be excluded from consideration as unphysical. It is interesting that in the case of inelastic scattering, going through the stage of a compound nucleus, this point cannot be excluded as unphysical, since after the interaction the neutron can be ejected from the compound nucleus with the same momentum vector that it had before the interaction. However, the situation is saved by the fact that the inelastic scattering reaction is threshold (as already indicated above in this section, its threshold is ~ hundreds of kiloelectronvolts) and its cross section at neutron energies corresponding to the zeroing of the energy decrement is zero.

Therefore, the obtained results demonstrate significant progress in understanding neutron retardation spectra in uranium-carbon and carbon media, particularly through the identification of a sharp maximum in the low-energy part of the spectrum, caused by the variation of the logarithmic energy decrement. The developed retardation theory allows for the consideration of the decrement's dependence on energy and temperature, unlike standard models, and also provides an explanation for the thermal interaction of neutrons with the nuclei of the retarding medium, which includes both energy loss and gain. The comparison with GEANT4 and the Monte Carlo method further confirms the reliability of the analytical expressions, opening up opportunities for their practical application. These results can be utilized for optimizing nuclear reactors, developing new retarding materials, enhancing radiation safety, and fostering innovations in small modular reactors, as well as for further experimental studies of neutron behavior in the low-energy range.

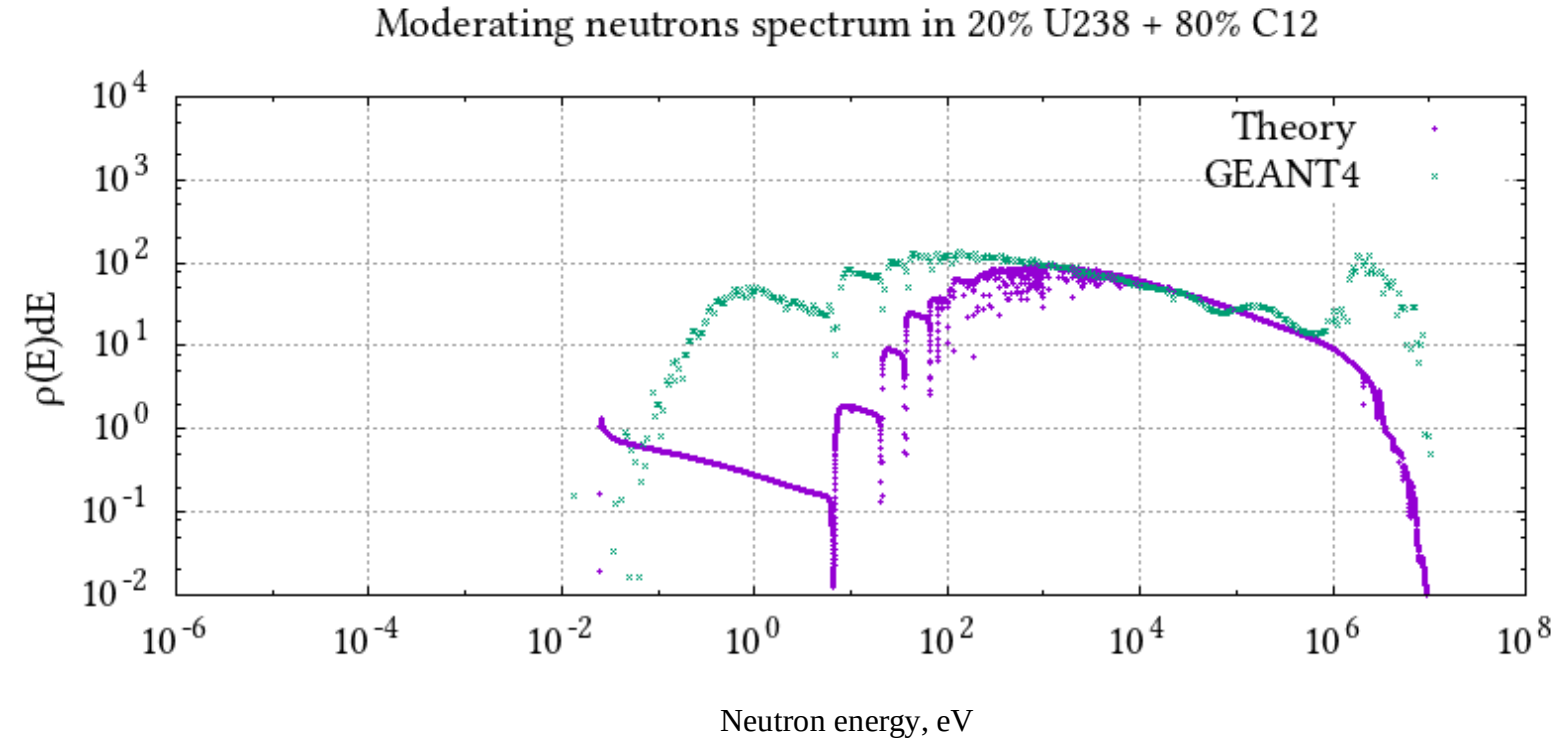

*Fig. 8.* Neutron energy spectrum for the uranium-carbon medium (20% uranium 238 and 80% carbon C12) calculated using expression (63) and the GEANT4 code with the fission neutron source given by expression (65) at the moderator temperature equal to 600 K

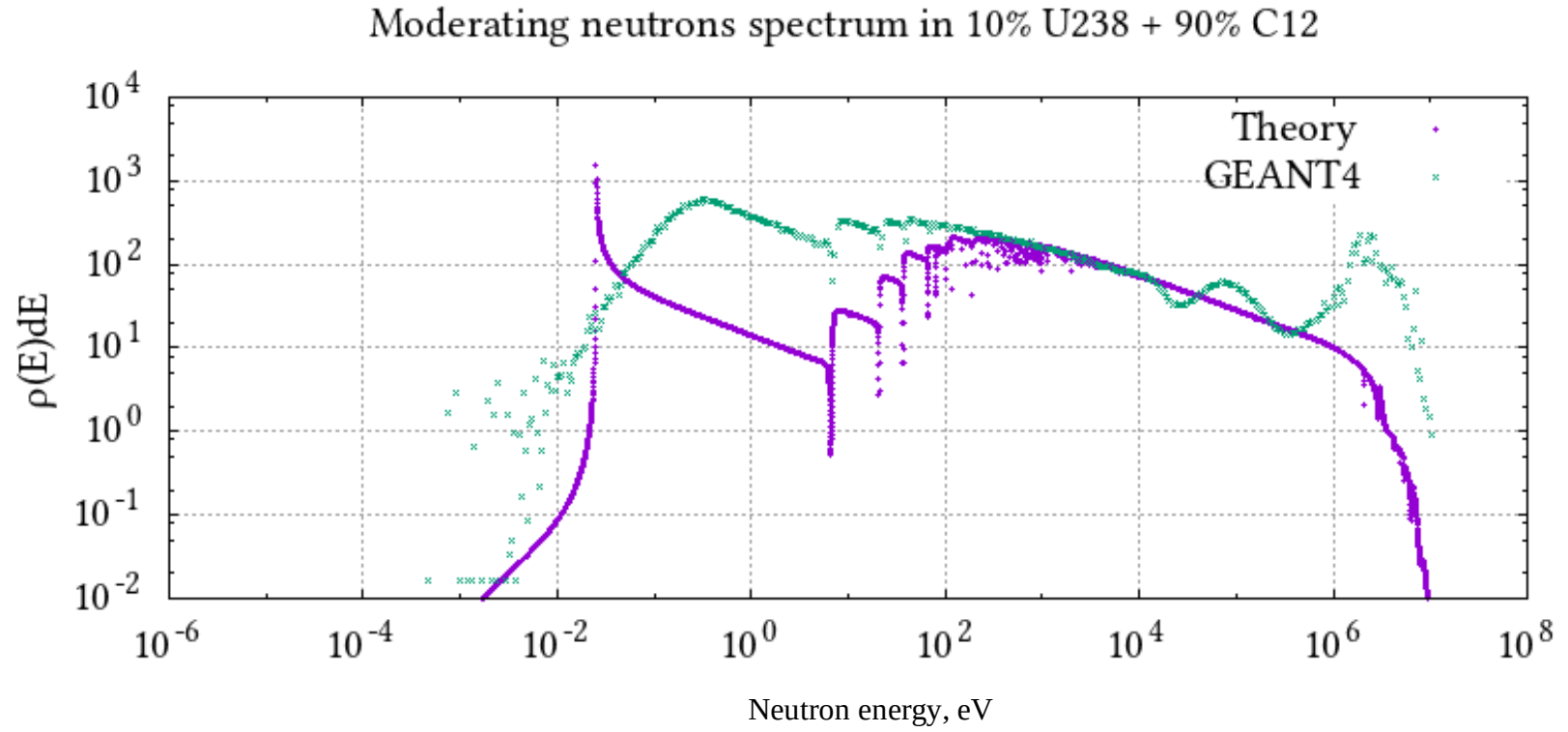

*Fig. 9.* Neutron energy spectrum for the uranium-carbon medium (10% uranium 238 and 90% carbon C12) calculated using expression (63) and the GEANT4 code with the fission neutron source given by expression (65) at the moderator temperature equal to 600 K.

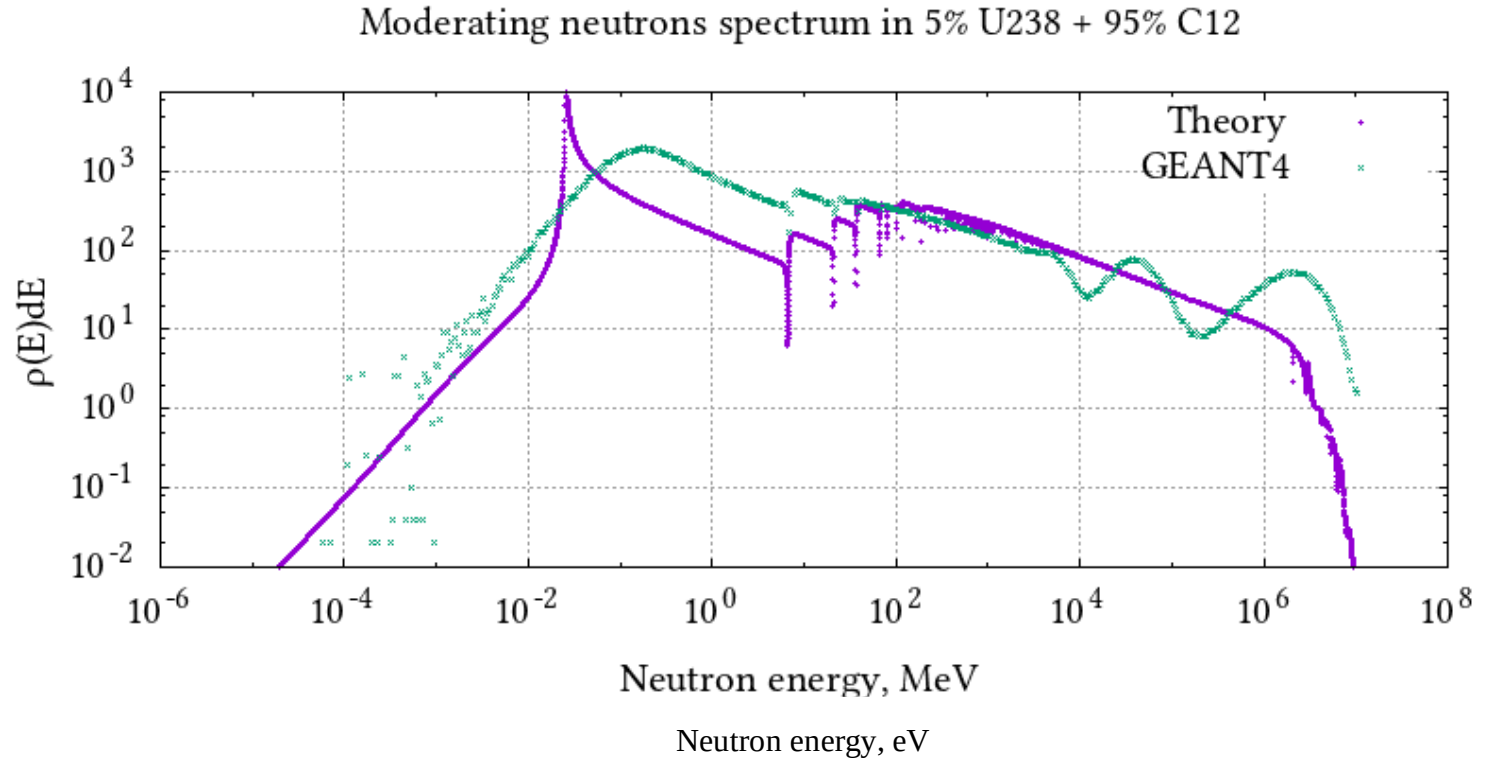

*Fig. 10.* Neutron energy spectrum for the uranium-carbon medium (5% uranium 238 and 95% carbon C12) calculated using expression (63) and the GEANT4 code with the fission neutron source given by expression (65) at the moderator temperature equal to 600 K.

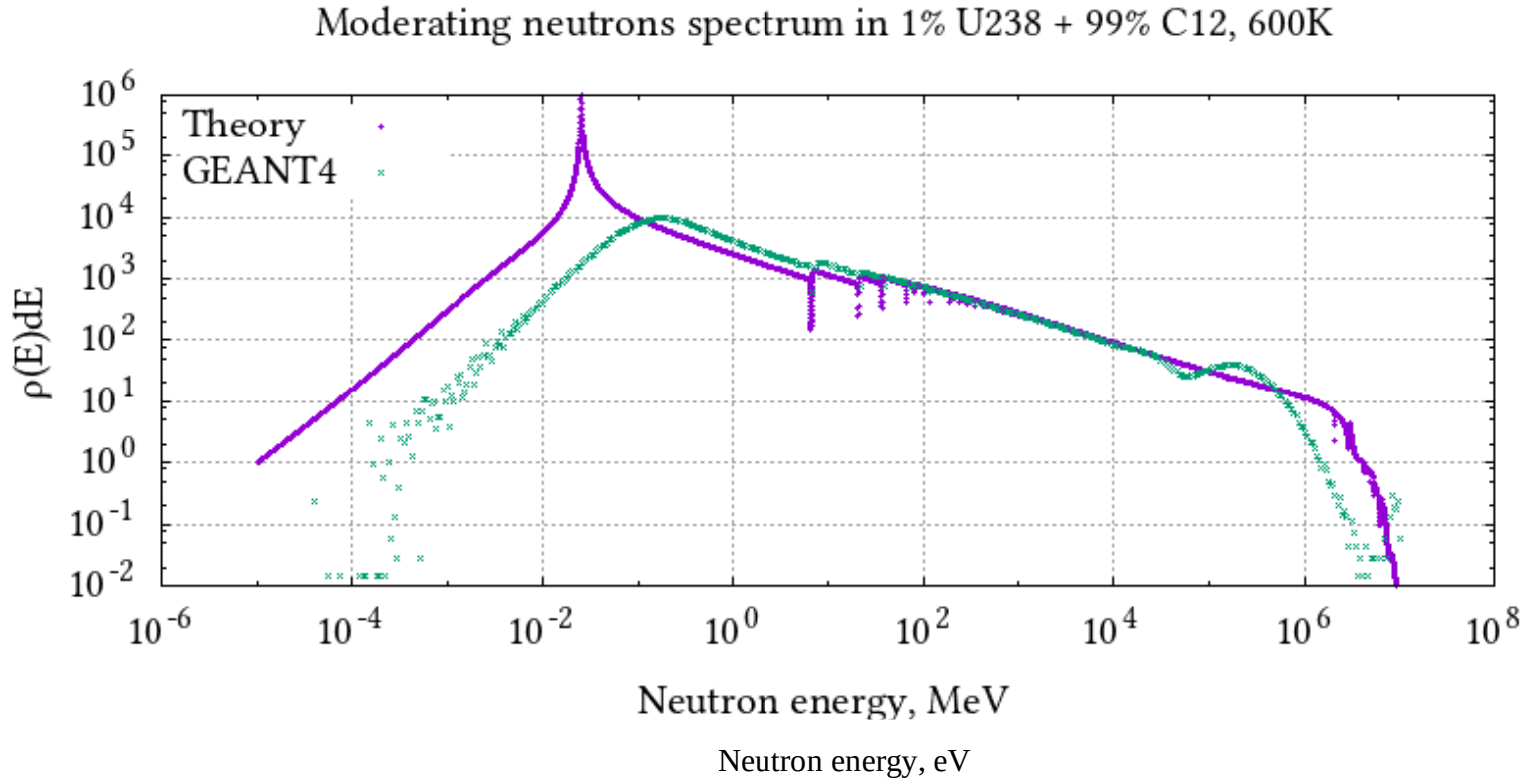

*Fig. 11.* Neutron energy spectrum for the uranium-carbon medium (1% uranium 238 and 99% carbon C12) calculated by expression (63) and using the GEANT4 code with the fission neutron source given by expression (65) at the moderator temperature equal to 600 K

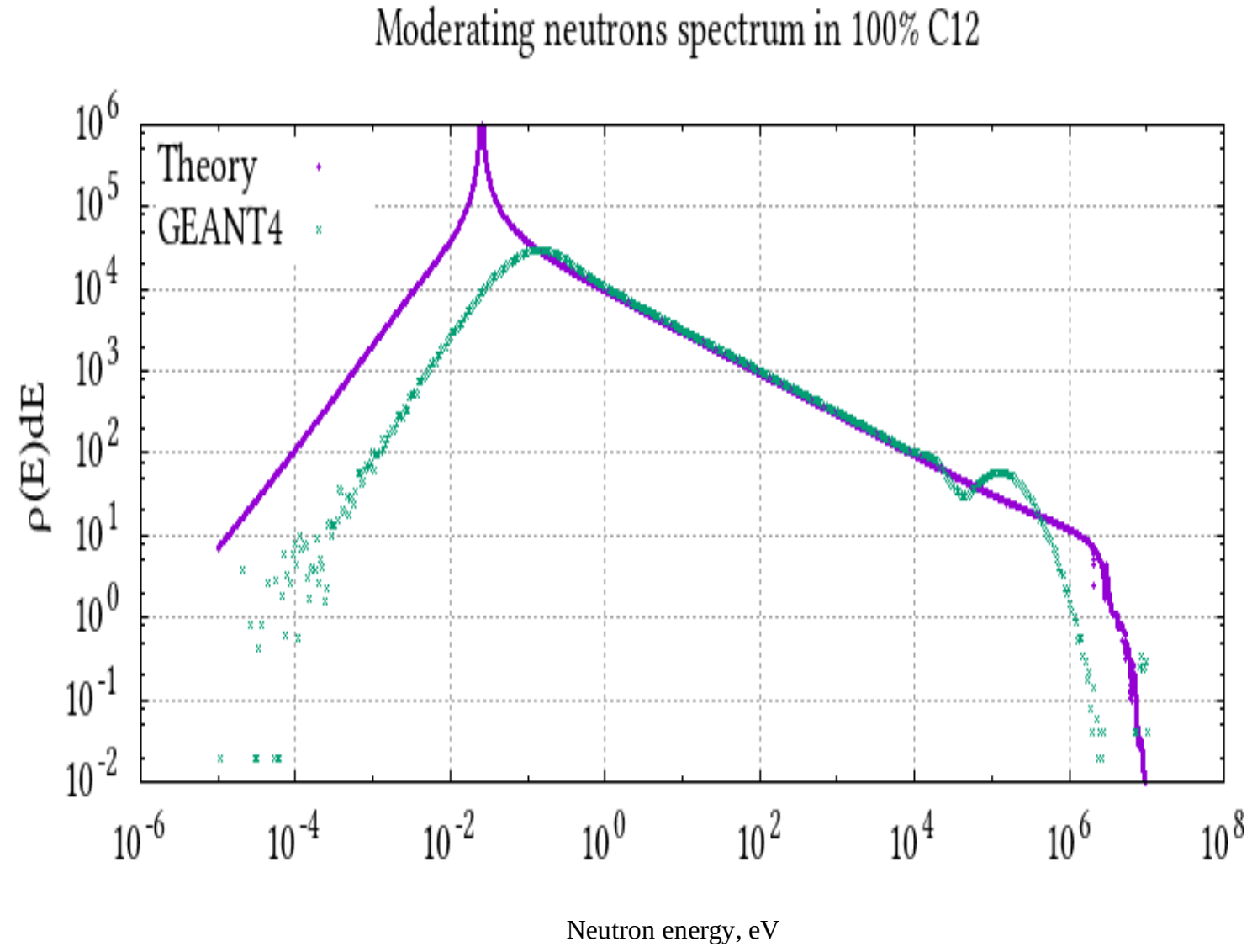

*Fig. 12.* Neutron energy spectrum for a carbon medium (100% carbon C12) calculated by expression (63) and using the GEANT4 code with the fission neutron source given by expression (65) at the moderator temperature equal to 600 K.

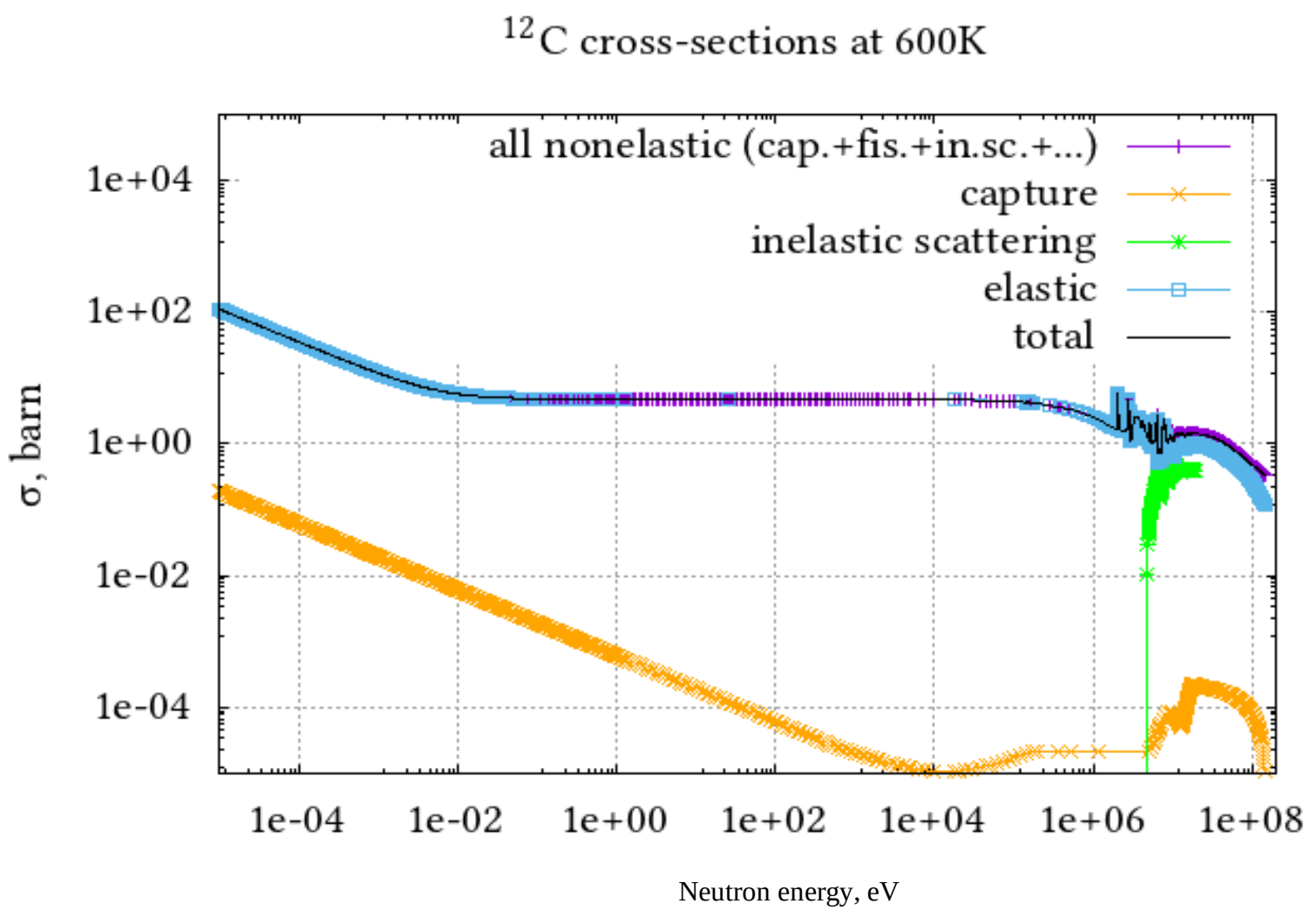

*Fig.13.* Deposits of neutron energy in micropertinents of neutron nuclear reactions for C12 carbon at temperatures above 600 K.

0

## Conclusions

In the scientific work, an analytical expression for the law of inelastic neutron scattering for an isotropic neutron source was obtained for the first time, including the temperature of the retarding medium as a parameter, as well as analytical expressions for the neutron flux density and the neutron retardation spectrum for an isotropic neutron source in retarding neutron-absorbing media (various reactor media with nuclei of several types), also dependent on temperature. These results open a new chapter in the interpretation of the physical nature of the processes that shape the type of neutron spectrum in the low-energy region, particularly revealing the influence of the behavior of elastic and inelastic neutron scattering cross-sections and the logarithmic energy decrement on the maximum of the neutron retardation spectrum. The nature of this maximum is linked to the retardation of a non-equilibrium neutron system on the thermalized system of nuclei in the retarding medium, rather than solely the thermodynamically equilibrium part described by the Maxwell distribution, marking a significant departure from traditional views.

Presented are graphs of energy spectra of neutrons slowing down in homogeneous hydrogen and uranium-carbon media, obtained through computer calculations based on the developed theory and the GEANT4 code. A comparative analysis of neutron spectra for uranium-carbon retarding media demonstrates good agreement between the calculations based on the developed retardation theory and those obtained using the Monte Carlo method, confirming the reliability of the derived analytical expressions. The significantly different behavior of elastic and inelastic neutron scattering cross-sections on the nuclei of various retarding media (ENDF/B-VII.0 (Agostinelli et al., 2003; Chadwick et al., 2011)) creates opportunities for experimental research into the influence of these cross-sections on the formation of the maximum of the neutron retardation spectrum in the low-energy region, as well as for the experimental verification of the developed models.

The practical significance of these results lies in the optimization of nuclear reactors through accurate prediction of neutron behavior, contributing to safer and more efficient designs, as well as in the development of retarding materials with optimal properties. They pave the way for experimental verification and standardization of neutron spectra, enhancing radiation safety by assessing dose loads, and foster innovations in energy, particularly in the creation of small modular reactors.

**Abstract (short)**

This research provides a new analytical framework for predicting how neutrons slow down and redistribute energy inside nuclear reactors under realistic temperature conditions. By explicitly accounting for inelastic neutron scattering and fuel temperature, the model improves the physical accuracy of neutron spectrum calculations that underpin reactor safety, efficiency, and accident analysis. These advances are particularly relevant for next-generation nuclear systems, including small modular reactors, fast reactors, and advanced fuel cycles, where temperature effects and non-equilibrium conditions are critical. More reliable neutron-kinetic models can support better regulatory decision-making, enhance safety margins, and reduce uncertainty in reactor licensing and risk assessment, contributing to safer and more efficient low-carbon energy systems.

**Funding**
This research received no external funding.

**Competing Interests**
The authors declare that they have no known competing financial interests or personal relationships that could have appeared to influence the work reported in this paper.

**Author Contributions**
Conceptualization: S.C., V.T., I.K.;
Methodology: S.C., V.T., V.V.;
Formal analysis: S.C., V.T.;
Investigation: S.C., V.T., V.V.;
Data curation: S.C., V.T.;
Writing—original draft preparation: S.C., V.T.;
Writing—review and editing: V.V., I.K.;
Visualization: V.V.;
Supervision: S.C.;
Project administration: I.K.

**Data Availability Statement**
The data supporting the findings of this study are publicly available. All nuclear data used in the calculations were obtained from the evaluated nuclear data library **ENDF/B-VII.1**, published in *Nuclear Data Sheets* (Vol. 112, pp. 2887–2996). The dataset is freely accessible via the National Nuclear Data Center (NNDC), Brookhaven National Laboratory: https://www.nndc.bnl.gov/endf/b7.1/
No new experimental data were generated in this study. All results can be reproduced using the cited data source and the methodological descriptions provided in the article.